\documentclass[aps,prl,twocolumn,longbibliography]{revtex4-1}

\usepackage{hyperref} 
\usepackage{graphicx}
\usepackage{bm}
\usepackage{amsmath}
\usepackage{amssymb}

\DeclareGraphicsExtensions{.png,.jpg,.eps}
\usepackage{xcolor}

\newcommand{\beq}{\begin{eqnarray}}
\newcommand{\eeq}{\end{eqnarray}}
\begin{document}

\author{Guangze Chen}
\affiliation{Institute for Theoretical Physics, ETH Zurich, 8093 Zurich, Switzerland}

\author{Wei Chen}
\email{pchenweis@gmail.com}
\affiliation{Institute for Theoretical Physics, ETH Zurich, 8093 Zurich, Switzerland}

\author{Oded Zilberberg}
\affiliation{Institute for Theoretical Physics, ETH Zurich, 8093 Zurich, Switzerland}

\title{Field-Effect Transistor based on Surface Negative Refraction in Weyl Nanowires}

\begin{abstract}
Weyl semimetals are characterized by their bulk Weyl points -- conical band touching points that carry a topological monopole charge -- and Fermi arc states that span between the Weyl points on the surface of the material.
Recently, significant progress has been made towards understanding
and measuring the physical properties of Weyl semimetals.
Yet, potential applications remain relatively sparse.
Here, we propose Weyl semimetal nanowires as field-effect transistors, dubbed WEYLFETs.
Specifically, applying gradient gate voltage along the nanowire, an electrical field is generated that effectively
tilts the open surfaces, thus, varying the relative orientation between
Fermi arcs on different surfaces. As a result, perfect negative refraction between
adjacent surfaces can occur and longitudinal conductance along the wire is suppressed. The WEYLFET offers a high on/off ratio with low power consumption. Adverse effects due to dispersive Fermi arcs and surface disorder are studied.
\end{abstract}

\date{\today}

\maketitle

Field-effect transistors (FETs) are electronic devices that use an electric field to control the flow of current through the device~\cite{galup2007mosfet}. There is a wide variety of materials and platforms used for various use cases of FETs; the majority thereof rely on semiconductor devices where the conduction channel can be switched off using an external gate. The conduction channel lies in the bulk of the semiconductor and early challenges in FET production concerned with surface passivation in order to overcome surface effects that prevented the gating from reaching the bulk~\cite{galup2007mosfet}. In parallel, low-power FETs' applications are in constant development where new materials and reduced dimensionality of the conduction channel play a crucial role~\cite{javey2003ballistic,schwierz2010graphene}.
Most recently, the progress of topological materials has opened a new avenue towards this goal based on the dissipationless chiral edge channels of the quantum anomalous Hall effect~\cite{Haldane88prl,Liu08prl,Yu10scn,hasan2010colloquium,qi2011topological,Chang13scn,Liu2016,chen19prb}.

Weyl semimetals are a class of 3D materials with conduction and valence bands that linearly touch at isolated points of the bulk spectrum~\cite{wan2011topological,murakami2007phase,Burkov11prl,wang2013three,Weng15prx,huang2015weyl}. The touching points are so-called Weyl points, around which the
electronic states can be effectively described by the Weyl equation~\cite{wan2011topological}.
Each Weyl point carries a monopole charge of Berry curvature, thus splitting momentum space into different regions of gapped spectra with different topology~\cite{wan2011topological,Yang11prb}. These unique bulk properties lead to electron chirality that offers  potential applications as a bulk photovoltaic effect~\cite{chan2017photocurrents,osterhoudt2019colossal}.

\begin{figure}
\center
\includegraphics[width=\linewidth]{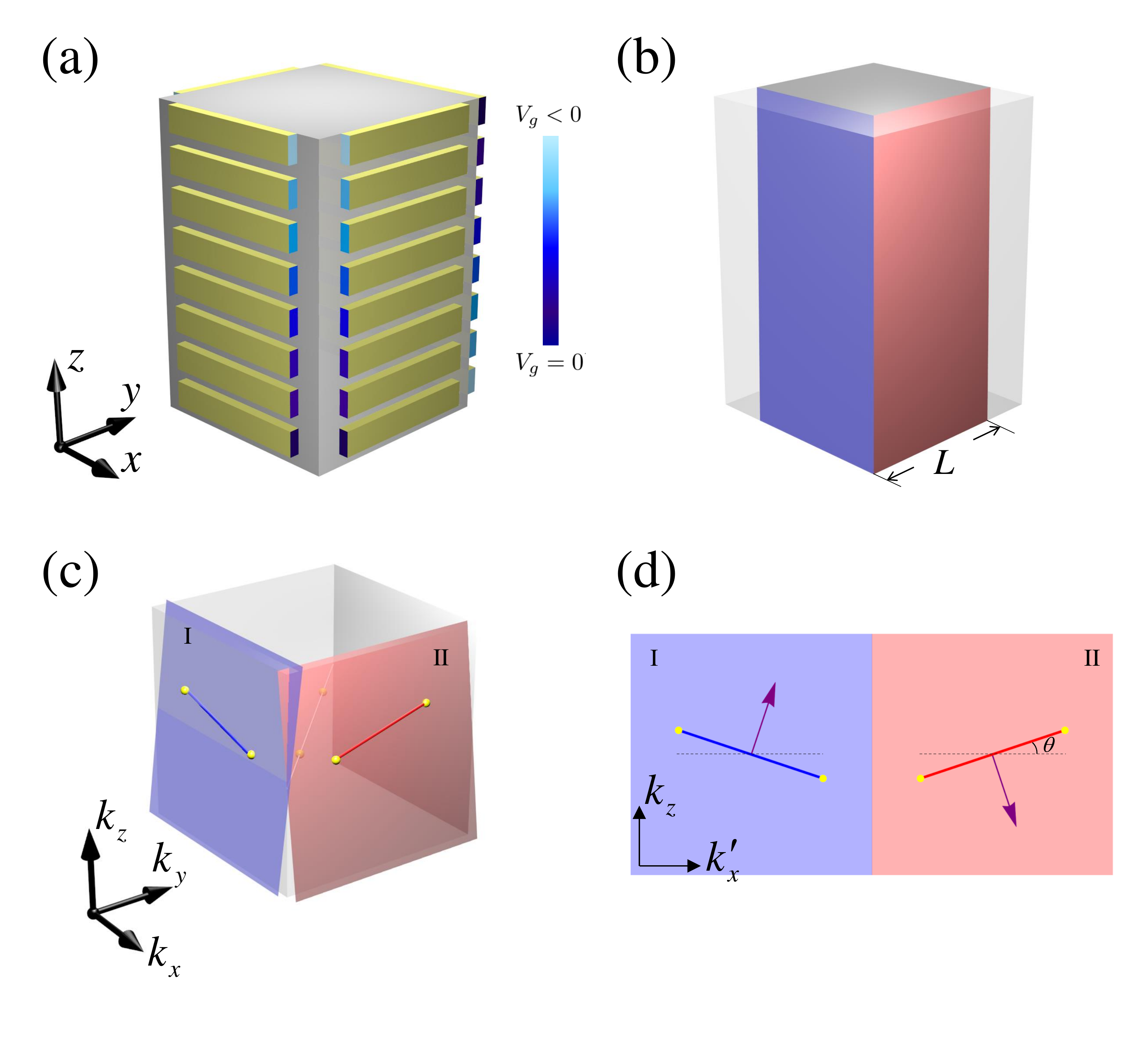}
\caption{Schematic of the Weyl semimetal field-effect transistor (WEYLFET). (a) The device is made of a Weyl semimetal nanowire (bulk core). Isolated metal gates are fabricated on top of the wire's surfaces with
opposite gate-voltage gradients on the front and back surfaces (mind the colorbar). (b) The gate voltage effectively results in tilting of the open surfaces. (c) The tilted surface Brillouin zones
lead to tilting of the Fermi arcs. (d) Two adjacent surfaces with opposite rotation
of Fermi arcs induce negative refraction.
}
\label{fig1}
\end{figure}

In parallel, the nontrivial bulk topology has a corresponding boundary effect in the form of Fermi arcs that appear in the surface Brillouin zone~\cite{wan2011topological}, which were recently observed experimentally~\cite{lv2015experimental,xu2015discovery,xu2015discovery2,xu2015experimental,xu2016observation,deng2016experimental,yang2015weyl,
huang2016spectroscopic,tamai2016fermi,jiang2017signature,belopolski2016discovery,lv2015observation}. These arcs have an open cut at the chemical potential, leading to directional transport on the surface. Furthermore, depending on the orientation of the surface, the Fermi arcs are projected from the bulk Weyl points differently~\cite{chen19arxiv}. This can lead to tunable surface  configurations where negative refraction may occur between different surfaces of the material~\cite{he2018topological}.

In this letter, we propose a new type of FET based on
field-controlled surface negative refraction in a Weyl nanowire,
or in short a WEYLFET.
We consider a Weyl semimetal nanowire covered by isolated metal gates on top of each surface, see Fig.~\ref{fig1}(a).
By imposing a slanted electric potential along the nanowire,
the redistribution of the electrons in the nanowire
adjusts the bearings of the open surfaces, see Fig. \ref{fig1}(b). For nanowires with properly chosen surface bearings relative to the orientation of the bulk Weyl nodes, the electric potential
results in an effective tilting of the Fermi arc in each surface Brillouin zone, see Fig. \ref{fig1}(c). Such relative tilting of Fermi arcs in the surface Brillouin zone can lead to perfect
negative refraction between adjacent surfaces~\cite{he2018topological,chen19arxiv}, see
Fig. \ref{fig1}(d). The surface negative refraction considerably suppresses the conduction
of electrons along the nanowire, i.e., for linear Fermi arcs, this effect produces a sharp
electrically-tunable on/off switch of the conductance. We analyze the adverse impact of dispersive Fermi
arcs, and the influence of surface disorder. Our proposal offers (i) a
robust surface effect that does not necessitate passivation,
(ii) reduced dissipation due to the backscattering-free channels, and
(iii) a tunable sharp on/off switch, all of which make our proposed
device a promising candidate for a low-power (WEYL-)FET.


In order to realize a WEYLFET with high on/off ratio,
it is essential to have ideal Weyl semimetals~\cite{Ruan16nc,Ruan16prl},
in which the Fermi energy crosses both Weyl points
and the transport is dominated by surface electrons.
Furthermore, we consider sufficiently small tilting angles, such that
reduction of the cross-section area of the nanowire is negligible.
Hence, the main effect of the slanted gate voltage is the tilting of the orientation of the Fermi arcs, and the WEYLFET is controlled by the modulated surface transport.

Instead of studying nanowires with various surface bearings, we focus on the bulk Weyl nanowire
with a general orientation of Weyl points,
while retaining the same termination configuration of the nanowire.
This approach is beneficial for the numerical study of the surface states~\cite{chen19arxiv}.
More concretely, we consider
the minimal inversion ($\mathcal{P}$)-symmetric
Weyl semimetal with two generally orientated Weyl points~\cite{chen2013specular}. We start with a Weyl semimetal with two Weyl points at $\pm(0,0,k_0)$:
\begin{eqnarray} \label{p}
\begin{split}
H(\bm{k})&=\hbar v(k_x\sigma_x+k_y\sigma_y)+M(k_0^2-\bm{k}^2)\sigma_z,
\end{split}
\end{eqnarray}
where $v$, $M$ and $k_0$ are parameters,
$\bm{k}=(k_x, k_y, k_z)$ is the wave vector, and $\sigma_{x,y,z}$ are Pauli
matrices acting on the pseudospin space.
The general orientation of Weyl points is achieved using the
rotational transformation $U$ to the effective
Hamiltonian $\mathcal{H}(\bm{k})= H(U^{-1}\bm{k})
$, e.g., the rotation by an angle $\varphi$ around the axis
$k_x=k_y,k_z=0$
yields two Weyl points located at $\pm k_0(-\frac{\sin\varphi}{\sqrt{2}},\frac{\sin\varphi}{\sqrt{2}},\cos\varphi)$.
We consider a nanowire along the $z$-direction with
a square cross section with side length $L$, cf.~Fig.~\ref{fig1}(b).
We take $\varphi_0=\pi/2$ as the starting point in the absence
of applied gate voltage $V$. The gating results in rotation of the Weyl points with angle $\varphi_0-\varphi(V)$. It corresponds to a tilting of the Fermi arcs by an angle $\theta(V)=\varphi_0-\tan^{-1}[\tan\varphi(V)/\sqrt{2}]$ in the
surface Brillouin zone~\cite{chen19arxiv} [Fig.~\ref{fig1}(d)].

\begin{figure}
\center
\includegraphics[width=\linewidth]{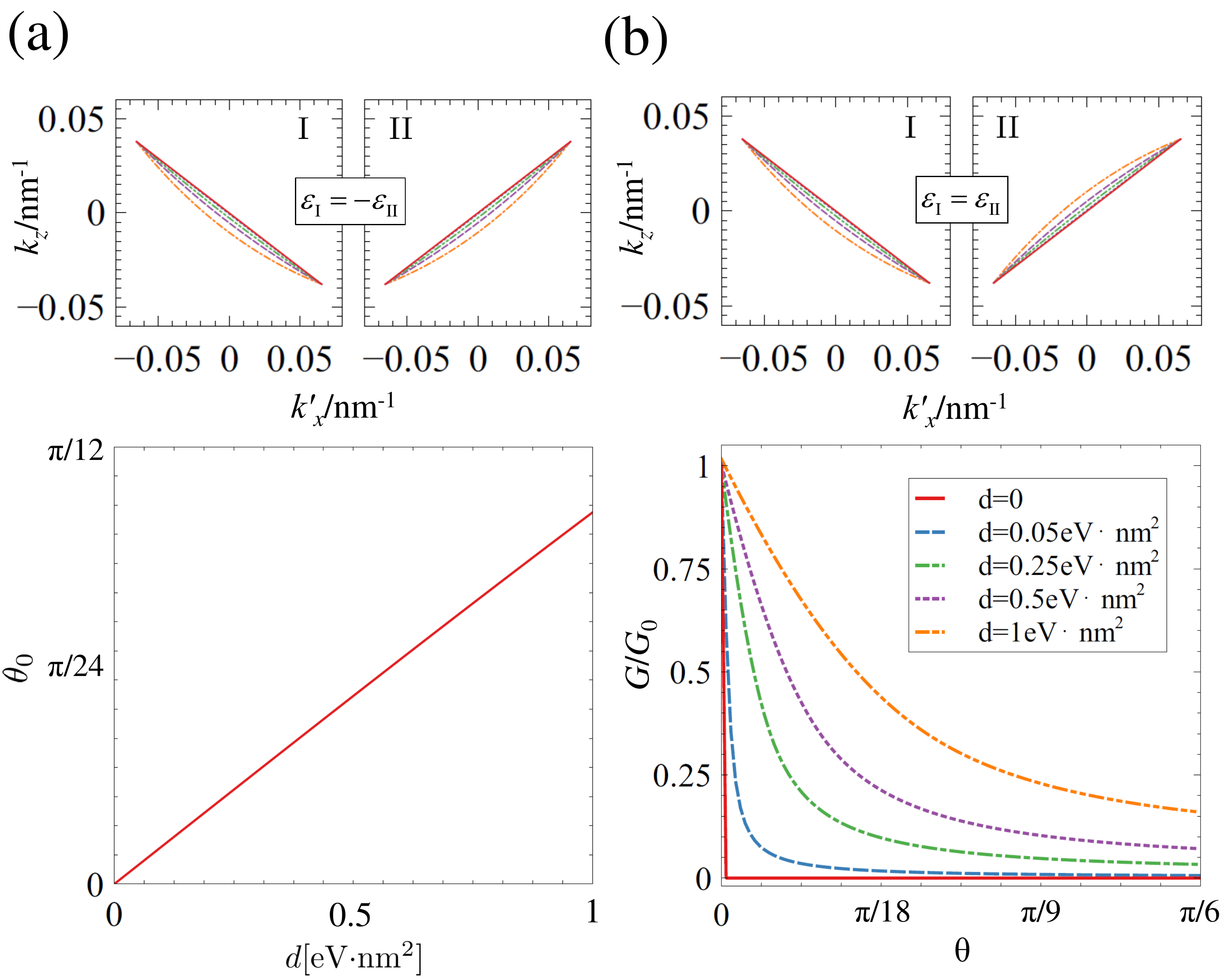}
\caption{(a) The
Fermi arcs defined by Eq. \eqref{surface_H} with different dispersion strength $d$ [labelled by the
legend in (b)] (upper panel) and the switch-off angle $\theta_0$ as a function of $d$ (lower panel), with $\varepsilon_{\text{I}}=-\varepsilon_{\text{II}}$. (b) The Fermi arcs defined by Eq. \eqref{surface_H} (upper panel) and the conductance $G$ [Eq.~\eqref{G}] as a function of $\theta$ (lower panel) with different dispersion $d$, and $\varepsilon_{\text{I}}=\varepsilon_{\text{II}}$. All results are obtained with parameters: $v_0=10^6$m/s and $k_0=0.1$nm$^{-1}$.
}
\label{fig2}
\end{figure}

For convenience, we unfold the four open surfaces
into the $x-z$ plane and label the longitudinal (transverse)
momentum by $k_z$ ($k'_x$), see Fig. \ref{fig1}(d).
The Fermi arcs on surfaces I and II can be
described by the effective Hamiltonian
\beq \label{surface_H}
H_{\text{I,II}}(\textbf{k})=\hbar v_0(\sin\theta k'_x\pm\cos\theta k_z)+\varepsilon_{\text{I,II}}(\textbf{k})\,,
\eeq
where $\textbf{k}=(k'_x,k_z)$ is the in-plane
momentum, $v_0$ is the velocity of the surface states,
and ``$\pm$" corresponds to surfaces I and II, respectively.
$\varepsilon_{\text{I,II}}$ is the dispersion term, which
introduces finite curvature to the Fermi arcs as in real materials.
When $\varepsilon_{\text{I,II}}=0$, the
straight Fermi arcs defined by $H_{\text{I,II}}=\sin\theta k'_x\pm\cos\theta k_z\equiv0$ have
 end points at $\pm k_0\left[-(\sin\varphi)/\sqrt{2},\cos\varphi\right]$
for surface I, and at
$\pm k_0\left[(\sin\varphi)/\sqrt{2},\cos\varphi\right]$ for surface II~\cite{chen19arxiv}.
The two Fermi arcs have opposite tilting, see Figs.~\ref{fig1}(c) and (d).
For a finite tilting angle ($\theta<\pi/2$), the velocity in the $z$-direction
is inverted as the electrons transfer through the boundary
between surfaces I and II,
leading to negative refraction, see Fig. \ref{fig1}(d).
This results in full suppression of electrons' flow
along the $z$-direction. This is the key mechanism behind the WEYLFET, i.e., even infinitesimal gating should lead to an on/off control of the electronic transport.

More realistically, Weyl semimetals exhibit dispersive Fermi arcs that depend on both bulk and surface details. Here, we consider a parabolic dispersion
$|\varepsilon_{\text{I,II}}|=d[k_0^2(1-\frac{1}{2}\sin^2\varphi)-{k'_x}^2-k_z^2]$, where the overall sign of the dispersive correction on each surface can change. By tuning the dispersion strength $\left|d\right|$, the Fermi
arcs become curved [cf.~Figs.~\ref{fig2}(a) and (b)], as observed in real materials.
Correspondingly, the electronic group velocities
$\bm{v}^{\text{I,II}}(\textbf{k})=(v^{\text{I,II}}_x,v^{\text{I,II}}_z)$  for the two
surface states become $\textbf{k}$-dependent. The zero-energy single-spin conductance along the wire can be evaluated quasi-classically by
\beq
\begin{split}\label{G}
G=2Le^2\rho_0\bar{v}_z, \ \ \
\bar{v}_z=\int_{-k_x^0}^{k_x^0}\frac{dk'_x}{2k_x^0}v_z(k'_x),
\end{split}
\eeq
with
\beq
\resizebox{.85\hsize}{!}{$\displaystyle v_z(k'_x)=\frac{v^{\text{I}}_z(k'_x,k_z)v^{\text{II}}_x(k'_{x2},k_z)+v^{\text{II}}_z(k'_{x2},k_z)v^{\text{I}}_x(k'_x,k_z)}{v^{\text{I}}_x(k'_x,k_z)+v^{\text{II}}_x(k'_{x2},k_z)},$}
\eeq
where the integral is taken over the incident states of surface I
and the velocity $\bm{v}^{\text{II}}(k'_{x2},k_z)$
is determined by the conservation of $k_z$ during the negative refraction.
In the above expressions, $\rho_0$ is the surface density of states per unit area,
$\bar{v}_z$ is the average velocity in the $z$-direction,
and $\pm k^0_x$ are the $k_x$ component of the Weyl points.

For the case of straight Fermi arcs ($d=0$)
with $\theta=0$, the conductance reduces to $G_0=2Le^2\rho_0v_0$.
As the Fermi arcs are tilted by a finite angle $\theta\neq0$,
we have $v^{\text{I}}_x=v^{\text{II}}_x$ and $v^{\text{I}}_z=-v^{\text{II}}_z$,
which results in perfect negative refraction [cf.~Fig.~\ref{fig1}(d)].
As a result, the conductance is completely
switched off by the gradient gate voltage, thus
realizing a WEYLFET.
This result also holds true when finite dispersion is added in the form of $\varepsilon_{\text{I}}=-\varepsilon_{\text{II}}$.
In this case, due to the existence of reflection,
the negative refraction is imperfect for small $\theta$,
and results in a finite switch-off angle $\theta_0$, see Fig \ref{fig2}(a)\cite{another_fn}.

We also investigate the case with equal dispersion,
i.e., $\varepsilon_{\text{I}}=\varepsilon_{\text{II}}$.
In this case, the conductance $G$ becomes a function of the tilting angle $\theta$, see Fig. \ref{fig2}(b)\cite{footnote}.
$G$ decreases with $\theta$,
indicating the suppression of transport by negative refraction. Whereas for straight Fermi arcs, $G$ exhibits a sharp switch-off, as the dispersion strength increases, the current
carried by refracted electrons cannot cancel out the incident current,
resulting in a net current, which reduces the on/off
ratio of the transistor for small $\theta$.

\begin{figure}
\center
\includegraphics[width=\linewidth]{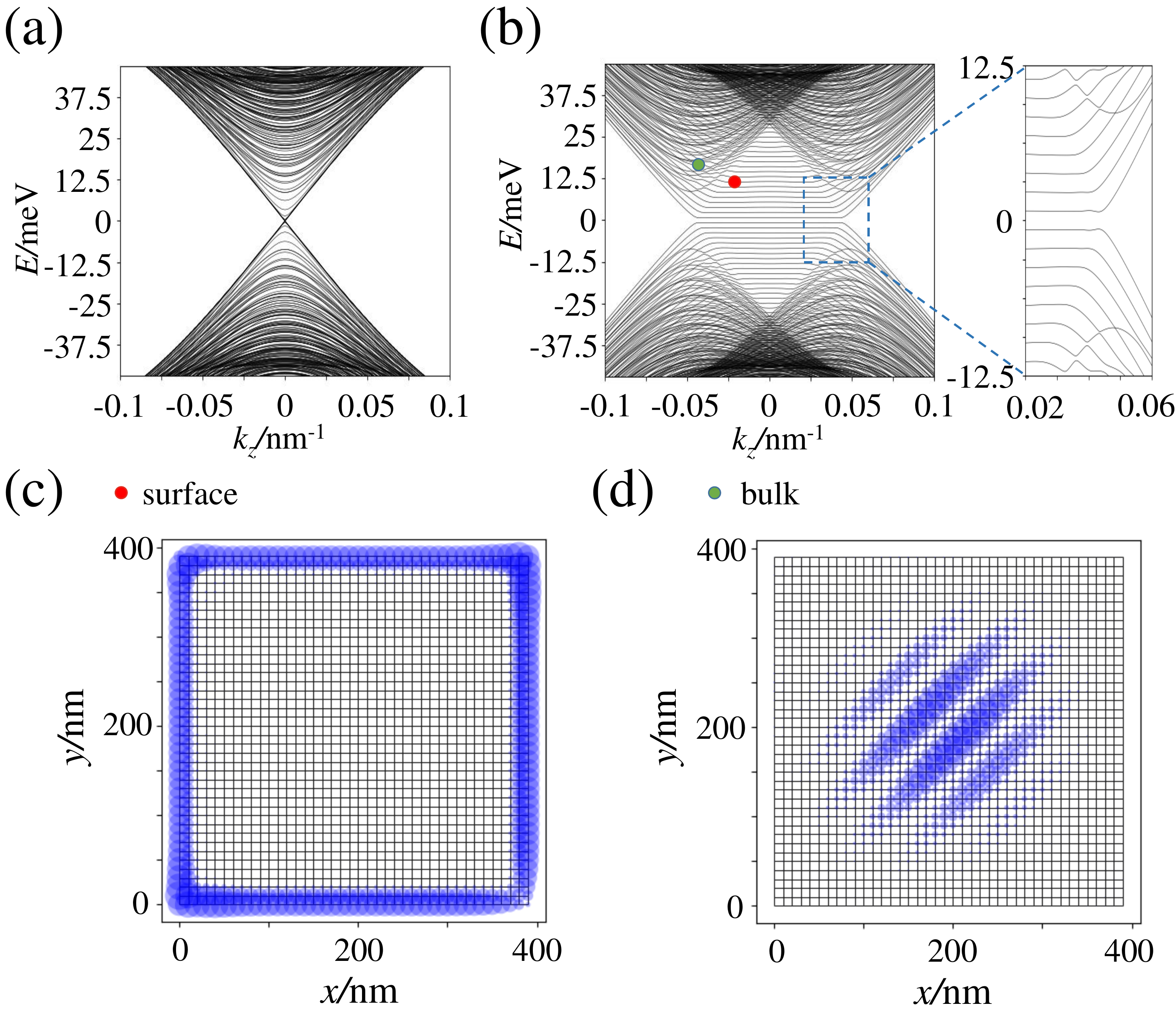}
\caption{Band structure of the Weyl nanowire for (a) $\theta=0$
and (b) $\theta=\pi/6$, numerically calculated with a cross section of 40$\times$40 sites.
Spatial distribution of (c) a surface state and (d) a bulk state.
Parameters used are $a=10$nm, $k_0=0.1$nm$^{-1}$, $M=4.375$eV$\cdot$nm$^2$ and $v=10^6$m/s.
}
\label{fig3}
\end{figure}

Complementary to the 2D surface negative refraction analysis above,
the on/off switch of the WEYLFET conductance can be
understood by studying the 1D band structure of the nanowire.
We map the Hamiltonian \eqref{p} onto a cubic lattice
through the substitutions
$
k_{i}\to a^{-1}\sin k_ia,\quad k_i^2\to2 a^{-2}(1-\cos k_ia)
$, with $a$ being the lattice constant. We can then solve the model and plot
its energy bands using KWANT~\cite{groth2014kwant}.
In Fig.~\ref{fig3}, we compare the resulting band structures
with and without the tilt of the Fermi arcs. For $\theta=0$,
the energy bands contributed by the surface states
are gapless, indicating a metallic phase. For finite
tilting angle $\theta$, a gap opening occurs for the surface states,
and the conduction is switched off for small bias voltages.
Therefore, under a small bias voltage,
the WEYLFET corresponds to a gate-tunable gap closing
and opening in the 1D picture, which is consistent
with the 2D picture of negative refraction, cf.~Fig.~\ref{fig1}(d).

\begin{figure}
\center
\includegraphics[width=\linewidth]{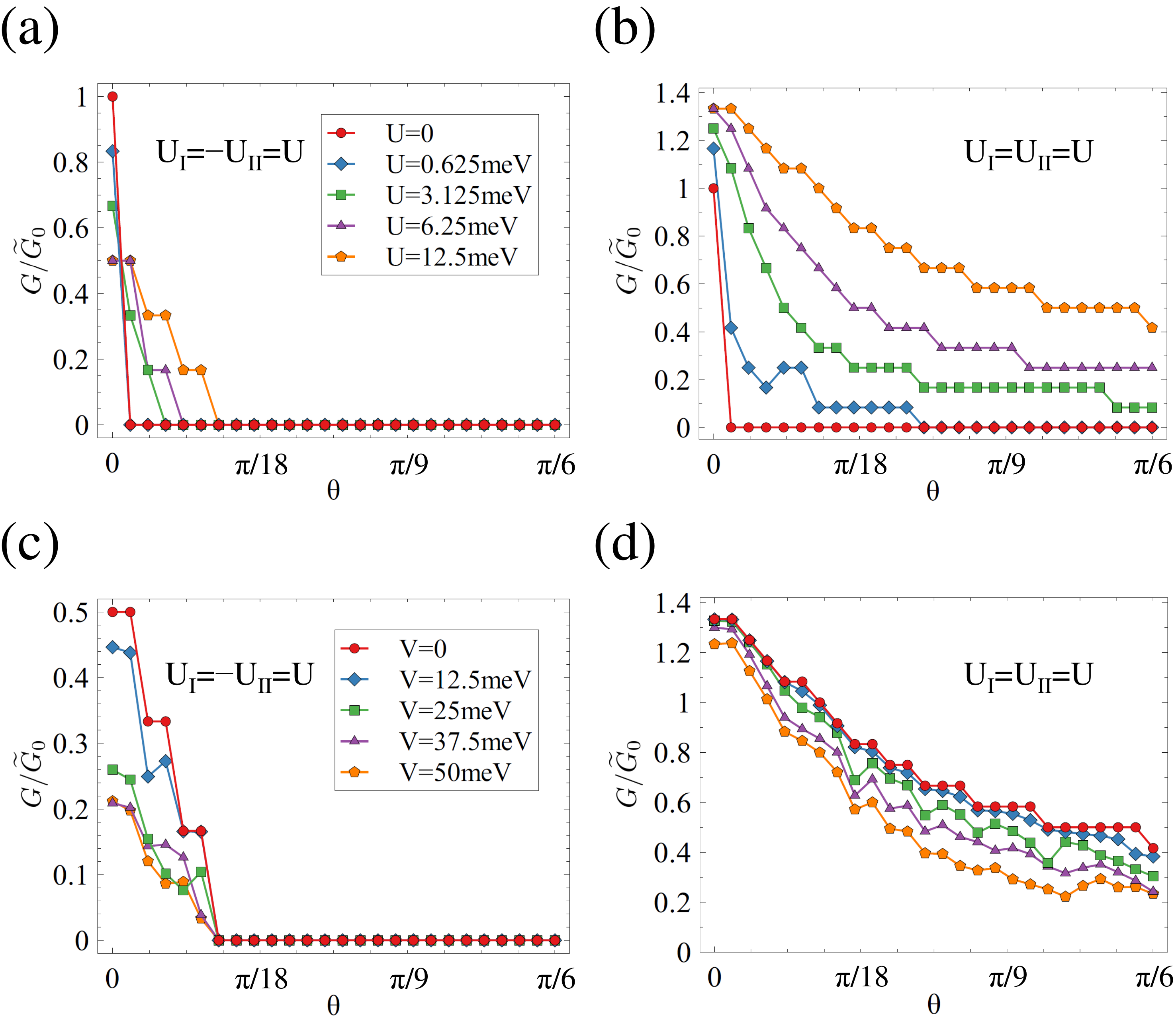}
\caption{Numerical simulation of adverse effect to the WEYLFET performance. (a) and (b): The conductance $G$ for different surface dispersions renormalized by $\widetilde{G}_0$ with $U=0, \theta=0$. (c) and (d): The conductance $G$ in the presence of surface disorder with $U=12.5$meV. All other parameters are the same as those in Fig.~\ref{fig3}.
}
\label{fig4}
\end{figure}

Next, we numerically calculate the conductance through the nanowire
 based on the cubic lattice model~\cite{groth2014kwant}.
We introduce surface dispersion
using an on-site potential $U$ on the surface layer;
because the surface states
have a $\textbf{k}$-dependent spatial
distribution perpendicular to the surface,
the on-site potential leads to a $\textbf{k}$-dependent
potential, or equivalently to an effective dispersion of the Fermi arcs.
The conductance along the nanowire as a function of
the tilting angle $\theta$ of the Fermi arcs is shown
in Figs.~\ref{fig4}(a) and (b).
In Fig.~\ref{fig4}(a), the onsite potential $U$ takes opposite values on adjacent surfaces.
Therefore, the dispersion on adjacent surfaces also takes opposite values, i.e.,
$\varepsilon_{\text{I}}=-\varepsilon_{\text{II}}$, and the on/off effect
is confirmed. The angle at which the conductance is
completely switched off increases with the onsite potential $U$, because curved Fermi arcs imply that reflection
processes exist in addition to negative refraction for small $\theta$.
Hence, in order to realize perfect negative refraction,
the tilting angle must exceed a critical value, cf.~Fig.~\ref{fig2}(a).
In Fig.~\ref{fig4}(b), the onsite potential $U$ takes the same value on all surfaces, and the
dispersions on the surfaces also become the same,
i.e., $\varepsilon_{\text{I}}=\varepsilon_{\text{II}}$. As a result, the
on/off ratio is reduced as $U$ increases, which is in
good agreement with the results in Fig.~\ref{fig2}(b).
Note that, here too, the surface
dispersion has an adverse effect on the WEYLFET performance, which reduces
the on/off ratio of the device. Therefore, in order to
make a WEYLFET with high efficiency, it is preferable to
use Weyl semimetals with weak and opposite surface dispersion.
Recent experiments have realized Fermi arc manipulation
by surface decoration~\cite{yang2019topological,morali2019fermi},
which paves the way to realize our proposal.

In real materials, surface roughness is unavoidable,
and we simulate this effect by including surface disorder.
For $\mathcal{P}$-symmetric Weyl semimetals,
the corresponding chiral surface states imply that electronic transport
is immune to surface disorder, i.e.,
the functionality of the WEYLFET should be robust to disorder.
In Figs.~\ref{fig4}(c) and (d), we present the calculated device conductance under different surface disorder strength
$V$ for both dispersion configurations ($U_\text{I}=\pm U_\text{II}$).
In both cases, the conductance decreases as surface disorder becomes
stronger. This stems from the enhancement of backscattering
that is induced by the disorder. However, the surface disorder
has no effect on the switch-off region in Fig.~\ref{fig4}(c),
where perfect negative refraction occurs.
In other words, the surface disorder has little effect
on the functionality of the WEYLFET and the on/off ratio
remains almost the same.

So far, we have solely analyzed a minimal $\mathcal{P}$-symmetric Weyl semimetal containing two Weyl points.
Onto this minimal model, we have introduced the general principle of gate tunable
Fermi arc tilting that can lead to negative refraction. Nevertheless,
further work will analyze even more realistic situations, including
(i) Weyl semimetals with multiple
pairs of Weyl points, as in most of the materials~\cite{lv2015experimental,
xu2015discovery,xu2015discovery2,
xu2015experimental,xu2016observation,deng2016experimental,yang2015weyl,
huang2016spectroscopic,tamai2016fermi,jiang2017signature,
belopolski2016discovery,lv2015observation}; here
the geometry of the nanowire should be properly chosen
such that the overlap between the projections of different Fermi arcs to the $z$-axis
is minimized. Otherwise, the reflection at
the boundary between different surfaces reduces the potency
of negative refraction, and accordingly the on/off ratio of the WEYLFET.
(ii) For time-reversal symmetric Weyl semimetals,
the Fermi arc states are not chiral, leading to enhanced
backscattering. In this case, we expect that surface disorder will have a stronger impact,
 and will reduce the on/off ratio of the WEYLFET.
Hence, for the time-reversal symmetric Weyl semimetal,
a nanowire with a clean surface is required
for a high-efficiency WEYLFET.
(iii) Another adverse effect can arise from
the deviation of the Fermi
level from the bulk Weyl points. In this case,
bulk electrons participate in the transport and contribute an overall
background to the conductance, which reduces the
on/off ratio of the WEYLFET. At the same time,
multiple scattering may occur in the bulk states,
which also increases the dissipation of the WEYLFET.
Furthermore, the finite bulk density of states
also brings considerable screening effect,
which reduces the efficiency of the slanted gate voltage.
Based on these observations, ideal Weyl semimetals are
needed for efficient WEYLFETs~\cite{Ruan16nc,Ruan16prl},
similar to their vital role in measuring other properties of Weyl semimetals.
(iv) Finally, in many Weyl semimetals,
the Fermi arcs' dispersion is complex,
making these surface states not much different from
2D normal metals, which cannot be used for WEYLFETs.
Consequently, the material candidate for the WEYLFET
should contain short Fermi arcs with small curvature~\cite{MnBi2Te4,wang19prb,soh2019ideal}.

In summary, we have explored a possible application of Weyl semimetal
as a field-effect transistor (WEYLFET) that is based on
electronic negative refraction between Fermi arcs on the surface of the device.
By using the gradient gate along the Weyl nanowire,
an on/off switch of the conductance can be achieved
with a high ratio. Ideal Weyl semimetals with vanishing
bulk density of states, chiral surface channels and small
Fermi arc curvature can
serve as good candidates for such high-efficiency WEYLFET
with low-power consumption.

\begin{acknowledgments}
We acknowledge financial support from the Swiss National Science
Foundation (SNSF) through Division II and the Careers Division.
\end{acknowledgments}


\begin{thebibliography}{45}%
\makeatletter
\providecommand \@ifxundefined [1]{%
 \@ifx{#1\undefined}
}%
\providecommand \@ifnum [1]{%
 \ifnum #1\expandafter \@firstoftwo
 \else \expandafter \@secondoftwo
 \fi
}%
\providecommand \@ifx [1]{%
 \ifx #1\expandafter \@firstoftwo
 \else \expandafter \@secondoftwo
 \fi
}%
\providecommand \natexlab [1]{#1}%
\providecommand \enquote  [1]{``#1''}%
\providecommand \bibnamefont  [1]{#1}%
\providecommand \bibfnamefont [1]{#1}%
\providecommand \citenamefont [1]{#1}%
\providecommand \href@noop [0]{\@secondoftwo}%
\providecommand \href [0]{\begingroup \@sanitize@url \@href}%
\providecommand \@href[1]{\@@startlink{#1}\@@href}%
\providecommand \@@href[1]{\endgroup#1\@@endlink}%
\providecommand \@sanitize@url [0]{\catcode `\\12\catcode `\$12\catcode
  `\&12\catcode `\#12\catcode `\^12\catcode `\_12\catcode `\%12\relax}%
\providecommand \@@startlink[1]{}%
\providecommand \@@endlink[0]{}%
\providecommand \url  [0]{\begingroup\@sanitize@url \@url }%
\providecommand \@url [1]{\endgroup\@href {#1}{\urlprefix }}%
\providecommand \urlprefix  [0]{URL }%
\providecommand \Eprint [0]{\href }%
\providecommand \doibase [0]{http://dx.doi.org/}%
\providecommand \selectlanguage [0]{\@gobble}%
\providecommand \bibinfo  [0]{\@secondoftwo}%
\providecommand \bibfield  [0]{\@secondoftwo}%
\providecommand \translation [1]{[#1]}%
\providecommand \BibitemOpen [0]{}%
\providecommand \bibitemStop [0]{}%
\providecommand \bibitemNoStop [0]{.\EOS\space}%
\providecommand \EOS [0]{\spacefactor3000\relax}%
\providecommand \BibitemShut  [1]{\csname bibitem#1\endcsname}%
\let\auto@bib@innerbib\@empty
\bibitem [{\citenamefont {Galup-Montoro}\ \emph {et~al.}(2007)\citenamefont
  {Galup-Montoro} \emph {et~al.}}]{galup2007mosfet}%
  \BibitemOpen
  \bibfield  {author} {\bibinfo {author} {\bibfnamefont {Carlos}\ \bibnamefont
  {Galup-Montoro}} \emph {et~al.},\ }\href@noop {} {\emph {\bibinfo {title}
  {MOSFET modeling for circuit analysis and design}}}\ (\bibinfo  {publisher}
  {World scientific},\ \bibinfo {year} {2007})\BibitemShut {NoStop}%
\bibitem [{\citenamefont {Javey}\ \emph {et~al.}(2003)\citenamefont {Javey},
  \citenamefont {Guo}, \citenamefont {Wang}, \citenamefont {Lundstrom},\ and\
  \citenamefont {Dai}}]{javey2003ballistic}%
  \BibitemOpen
  \bibfield  {author} {\bibinfo {author} {\bibfnamefont {Ali}\ \bibnamefont
  {Javey}}, \bibinfo {author} {\bibfnamefont {Jing}\ \bibnamefont {Guo}},
  \bibinfo {author} {\bibfnamefont {Qian}\ \bibnamefont {Wang}}, \bibinfo
  {author} {\bibfnamefont {Mark}\ \bibnamefont {Lundstrom}}, \ and\ \bibinfo
  {author} {\bibfnamefont {Hongjie}\ \bibnamefont {Dai}},\ }\bibfield  {title}
  {\enquote {\bibinfo {title} {Ballistic carbon nanotube field-effect
  transistors},}\ }\href@noop {} {\bibfield  {journal} {\bibinfo  {journal}
  {nature}\ }\textbf {\bibinfo {volume} {424}},\ \bibinfo {pages} {654}
  (\bibinfo {year} {2003})}\BibitemShut {NoStop}%
\bibitem [{\citenamefont {Schwierz}(2010)}]{schwierz2010graphene}%
  \BibitemOpen
  \bibfield  {author} {\bibinfo {author} {\bibfnamefont {Frank}\ \bibnamefont
  {Schwierz}},\ }\bibfield  {title} {\enquote {\bibinfo {title} {Graphene
  transistors},}\ }\href@noop {} {\bibfield  {journal} {\bibinfo  {journal}
  {Nature nanotechnology}\ }\textbf {\bibinfo {volume} {5}},\ \bibinfo {pages}
  {487} (\bibinfo {year} {2010})}\BibitemShut {NoStop}%
\bibitem [{\citenamefont {Haldane}(1988)}]{Haldane88prl}%
  \BibitemOpen
  \bibfield  {author} {\bibinfo {author} {\bibfnamefont {F.~D.~M.}\
  \bibnamefont {Haldane}},\ }\bibfield  {title} {\enquote {\bibinfo {title}
  {Model for a quantum hall effect without landau levels: Condensed-matter
  realization of the "parity anomaly"},}\ }\href {\doibase
  10.1103/PhysRevLett.61.2015} {\bibfield  {journal} {\bibinfo  {journal}
  {Phys. Rev. Lett.}\ }\textbf {\bibinfo {volume} {61}},\ \bibinfo {pages}
  {2015--2018} (\bibinfo {year} {1988})}\BibitemShut {NoStop}%
\bibitem [{\citenamefont {Liu}\ \emph {et~al.}(2008)\citenamefont {Liu},
  \citenamefont {Qi}, \citenamefont {Dai}, \citenamefont {Fang},\ and\
  \citenamefont {Zhang}}]{Liu08prl}%
  \BibitemOpen
  \bibfield  {author} {\bibinfo {author} {\bibfnamefont {Chao-Xing}\
  \bibnamefont {Liu}}, \bibinfo {author} {\bibfnamefont {Xiao-Liang}\
  \bibnamefont {Qi}}, \bibinfo {author} {\bibfnamefont {Xi}~\bibnamefont
  {Dai}}, \bibinfo {author} {\bibfnamefont {Zhong}\ \bibnamefont {Fang}}, \
  and\ \bibinfo {author} {\bibfnamefont {Shou-Cheng}\ \bibnamefont {Zhang}},\
  }\bibfield  {title} {\enquote {\bibinfo {title} {Quantum anomalous hall
  effect in ${\mathrm{hg}}_{1\ensuremath{-}y}{\mathrm{mn}}_{y}\mathrm{Te}$
  quantum wells},}\ }\href {\doibase 10.1103/PhysRevLett.101.146802} {\bibfield
   {journal} {\bibinfo  {journal} {Phys. Rev. Lett.}\ }\textbf {\bibinfo
  {volume} {101}},\ \bibinfo {pages} {146802} (\bibinfo {year}
  {2008})}\BibitemShut {NoStop}%
\bibitem [{\citenamefont {Yu}\ \emph {et~al.}(2010)\citenamefont {Yu},
  \citenamefont {Zhang}, \citenamefont {Zhang}, \citenamefont {Zhang},
  \citenamefont {Dai},\ and\ \citenamefont {Fang}}]{Yu10scn}%
  \BibitemOpen
  \bibfield  {author} {\bibinfo {author} {\bibfnamefont {R.}~\bibnamefont
  {Yu}}, \bibinfo {author} {\bibfnamefont {W.}~\bibnamefont {Zhang}}, \bibinfo
  {author} {\bibfnamefont {H.-J.}\ \bibnamefont {Zhang}}, \bibinfo {author}
  {\bibfnamefont {S.-C.}\ \bibnamefont {Zhang}}, \bibinfo {author}
  {\bibfnamefont {X.}~\bibnamefont {Dai}}, \ and\ \bibinfo {author}
  {\bibfnamefont {Z.}~\bibnamefont {Fang}},\ }\bibfield  {title} {\enquote
  {\bibinfo {title} {Quantized anomalous hall effect in magnetic topological
  insulators},}\ }\href {\doibase 10.1126/science.1187485} {\bibfield
  {journal} {\bibinfo  {journal} {Science}\ }\textbf {\bibinfo {volume}
  {329}},\ \bibinfo {pages} {61--64} (\bibinfo {year} {2010})}\BibitemShut
  {NoStop}%
\bibitem [{\citenamefont {Hasan}\ and\ \citenamefont
  {Kane}(2010)}]{hasan2010colloquium}%
  \BibitemOpen
  \bibfield  {author} {\bibinfo {author} {\bibfnamefont {M~Zahid}\ \bibnamefont
  {Hasan}}\ and\ \bibinfo {author} {\bibfnamefont {Charles~L}\ \bibnamefont
  {Kane}},\ }\bibfield  {title} {\enquote {\bibinfo {title} {Colloquium:
  topological insulators},}\ }\href@noop {} {\bibfield  {journal} {\bibinfo
  {journal} {Reviews of modern physics}\ }\textbf {\bibinfo {volume} {82}},\
  \bibinfo {pages} {3045} (\bibinfo {year} {2010})}\BibitemShut {NoStop}%
\bibitem [{\citenamefont {Qi}\ and\ \citenamefont
  {Zhang}(2011)}]{qi2011topological}%
  \BibitemOpen
  \bibfield  {author} {\bibinfo {author} {\bibfnamefont {Xiao-Liang}\
  \bibnamefont {Qi}}\ and\ \bibinfo {author} {\bibfnamefont {Shou-Cheng}\
  \bibnamefont {Zhang}},\ }\bibfield  {title} {\enquote {\bibinfo {title}
  {Topological insulators and superconductors},}\ }\href@noop {} {\bibfield
  {journal} {\bibinfo  {journal} {Reviews of Modern Physics}\ }\textbf
  {\bibinfo {volume} {83}},\ \bibinfo {pages} {1057} (\bibinfo {year}
  {2011})}\BibitemShut {NoStop}%
\bibitem [{\citenamefont {Chang}\ \emph {et~al.}(2013)\citenamefont {Chang},
  \citenamefont {Zhang}, \citenamefont {Feng}, \citenamefont {Shen},
  \citenamefont {Zhang}, \citenamefont {Guo}, \citenamefont {Li}, \citenamefont
  {Ou}, \citenamefont {Wei}, \citenamefont {Wang}, \citenamefont {Ji},
  \citenamefont {Feng}, \citenamefont {Ji}, \citenamefont {Chen}, \citenamefont
  {Jia}, \citenamefont {Dai}, \citenamefont {Fang}, \citenamefont {Zhang},
  \citenamefont {He}, \citenamefont {Wang}, \citenamefont {Lu}, \citenamefont
  {Ma},\ and\ \citenamefont {Xue}}]{Chang13scn}%
  \BibitemOpen
  \bibfield  {author} {\bibinfo {author} {\bibfnamefont {C.-Z.}\ \bibnamefont
  {Chang}}, \bibinfo {author} {\bibfnamefont {J.}~\bibnamefont {Zhang}},
  \bibinfo {author} {\bibfnamefont {X.}~\bibnamefont {Feng}}, \bibinfo {author}
  {\bibfnamefont {J.}~\bibnamefont {Shen}}, \bibinfo {author} {\bibfnamefont
  {Z.}~\bibnamefont {Zhang}}, \bibinfo {author} {\bibfnamefont
  {M.}~\bibnamefont {Guo}}, \bibinfo {author} {\bibfnamefont {K.}~\bibnamefont
  {Li}}, \bibinfo {author} {\bibfnamefont {Y.}~\bibnamefont {Ou}}, \bibinfo
  {author} {\bibfnamefont {P.}~\bibnamefont {Wei}}, \bibinfo {author}
  {\bibfnamefont {L.-L.}\ \bibnamefont {Wang}}, \bibinfo {author}
  {\bibfnamefont {Z.-Q.}\ \bibnamefont {Ji}}, \bibinfo {author} {\bibfnamefont
  {Y.}~\bibnamefont {Feng}}, \bibinfo {author} {\bibfnamefont {S.}~\bibnamefont
  {Ji}}, \bibinfo {author} {\bibfnamefont {X.}~\bibnamefont {Chen}}, \bibinfo
  {author} {\bibfnamefont {J.}~\bibnamefont {Jia}}, \bibinfo {author}
  {\bibfnamefont {X.}~\bibnamefont {Dai}}, \bibinfo {author} {\bibfnamefont
  {Z.}~\bibnamefont {Fang}}, \bibinfo {author} {\bibfnamefont {S.-C.}\
  \bibnamefont {Zhang}}, \bibinfo {author} {\bibfnamefont {K.}~\bibnamefont
  {He}}, \bibinfo {author} {\bibfnamefont {Y.}~\bibnamefont {Wang}}, \bibinfo
  {author} {\bibfnamefont {L.}~\bibnamefont {Lu}}, \bibinfo {author}
  {\bibfnamefont {X.-C.}\ \bibnamefont {Ma}}, \ and\ \bibinfo {author}
  {\bibfnamefont {Q.-K.}\ \bibnamefont {Xue}},\ }\bibfield  {title} {\enquote
  {\bibinfo {title} {Experimental observation of the quantum anomalous hall
  effect in a magnetic topological insulator},}\ }\href {\doibase
  10.1126/science.1234414} {\bibfield  {journal} {\bibinfo  {journal}
  {Science}\ }\textbf {\bibinfo {volume} {340}},\ \bibinfo {pages} {167--170}
  (\bibinfo {year} {2013})}\BibitemShut {NoStop}%
\bibitem [{\citenamefont {Liu}\ \emph {et~al.}(2016)\citenamefont {Liu},
  \citenamefont {Zhang},\ and\ \citenamefont {Qi}}]{Liu2016}%
  \BibitemOpen
  \bibfield  {author} {\bibinfo {author} {\bibfnamefont {Chao-Xing}\
  \bibnamefont {Liu}}, \bibinfo {author} {\bibfnamefont {Shou-Cheng}\
  \bibnamefont {Zhang}}, \ and\ \bibinfo {author} {\bibfnamefont {Xiao-Liang}\
  \bibnamefont {Qi}},\ }\bibfield  {title} {\enquote {\bibinfo {title} {The
  quantum anomalous hall effect: Theory and experiment},}\ }\href {\doibase
  10.1146/annurev-conmatphys-031115-011417} {\bibfield  {journal} {\bibinfo
  {journal} {Annual Review of Condensed Matter Physics}\ }\textbf {\bibinfo
  {volume} {7}},\ \bibinfo {pages} {301--321} (\bibinfo {year}
  {2016})}\BibitemShut {NoStop}%
\bibitem [{\citenamefont {Chen}\ \emph
  {et~al.}(2019{\natexlab{a}})\citenamefont {Chen}, \citenamefont {Chen},
  \citenamefont {Shao},\ and\ \citenamefont {Xing}}]{chen19prb}%
  \BibitemOpen
  \bibfield  {author} {\bibinfo {author} {\bibfnamefont {Xi-Rong}\ \bibnamefont
  {Chen}}, \bibinfo {author} {\bibfnamefont {Wei}\ \bibnamefont {Chen}},
  \bibinfo {author} {\bibfnamefont {L.~B.}\ \bibnamefont {Shao}}, \ and\
  \bibinfo {author} {\bibfnamefont {D.~Y.}\ \bibnamefont {Xing}},\ }\bibfield
  {title} {\enquote {\bibinfo {title} {Engineering chiral edge states in
  two-dimensional topological insulator/ferromagnetic insulator
  heterostructures},}\ }\href {\doibase 10.1103/PhysRevB.99.085417} {\bibfield
  {journal} {\bibinfo  {journal} {Phys. Rev. B}\ }\textbf {\bibinfo {volume}
  {99}},\ \bibinfo {pages} {085417} (\bibinfo {year}
  {2019}{\natexlab{a}})}\BibitemShut {NoStop}%
\bibitem [{\citenamefont {Wan}\ \emph {et~al.}(2011)\citenamefont {Wan},
  \citenamefont {Turner}, \citenamefont {Vishwanath},\ and\ \citenamefont
  {Savrasov}}]{wan2011topological}%
  \BibitemOpen
  \bibfield  {author} {\bibinfo {author} {\bibfnamefont {Xiangang}\
  \bibnamefont {Wan}}, \bibinfo {author} {\bibfnamefont {Ari~M}\ \bibnamefont
  {Turner}}, \bibinfo {author} {\bibfnamefont {Ashvin}\ \bibnamefont
  {Vishwanath}}, \ and\ \bibinfo {author} {\bibfnamefont {Sergey~Y}\
  \bibnamefont {Savrasov}},\ }\bibfield  {title} {\enquote {\bibinfo {title}
  {Topological semimetal and fermi-arc surface states in the electronic
  structure of pyrochlore iridates},}\ }\href@noop {} {\bibfield  {journal}
  {\bibinfo  {journal} {Physical Review B}\ }\textbf {\bibinfo {volume} {83}},\
  \bibinfo {pages} {205101} (\bibinfo {year} {2011})}\BibitemShut {NoStop}%
\bibitem [{\citenamefont {Murakami}(2007)}]{murakami2007phase}%
  \BibitemOpen
  \bibfield  {author} {\bibinfo {author} {\bibfnamefont {Shuichi}\ \bibnamefont
  {Murakami}},\ }\bibfield  {title} {\enquote {\bibinfo {title} {Phase
  transition between the quantum spin hall and insulator phases in 3d:
  emergence of a topological gapless phase},}\ }\href@noop {} {\bibfield
  {journal} {\bibinfo  {journal} {New Journal of Physics}\ }\textbf {\bibinfo
  {volume} {9}},\ \bibinfo {pages} {356} (\bibinfo {year} {2007})}\BibitemShut
  {NoStop}%
\bibitem [{\citenamefont {Burkov}\ and\ \citenamefont
  {Balents}(2011)}]{Burkov11prl}%
  \BibitemOpen
  \bibfield  {author} {\bibinfo {author} {\bibfnamefont {A.~A.}\ \bibnamefont
  {Burkov}}\ and\ \bibinfo {author} {\bibfnamefont {Leon}\ \bibnamefont
  {Balents}},\ }\bibfield  {title} {\enquote {\bibinfo {title} {Weyl semimetal
  in a topological insulator multilayer},}\ }\href {\doibase
  10.1103/PhysRevLett.107.127205} {\bibfield  {journal} {\bibinfo  {journal}
  {Phys. Rev. Lett.}\ }\textbf {\bibinfo {volume} {107}},\ \bibinfo {pages}
  {127205} (\bibinfo {year} {2011})}\BibitemShut {NoStop}%
\bibitem [{\citenamefont {Wang}\ \emph {et~al.}(2013)\citenamefont {Wang},
  \citenamefont {Weng}, \citenamefont {Wu}, \citenamefont {Dai},\ and\
  \citenamefont {Fang}}]{wang2013three}%
  \BibitemOpen
  \bibfield  {author} {\bibinfo {author} {\bibfnamefont {Zhijun}\ \bibnamefont
  {Wang}}, \bibinfo {author} {\bibfnamefont {Hongming}\ \bibnamefont {Weng}},
  \bibinfo {author} {\bibfnamefont {Quansheng}\ \bibnamefont {Wu}}, \bibinfo
  {author} {\bibfnamefont {Xi}~\bibnamefont {Dai}}, \ and\ \bibinfo {author}
  {\bibfnamefont {Zhong}\ \bibnamefont {Fang}},\ }\bibfield  {title} {\enquote
  {\bibinfo {title} {Three-dimensional dirac semimetal and quantum transport in
  cd 3 as 2},}\ }\href@noop {} {\bibfield  {journal} {\bibinfo  {journal}
  {Physical Review B}\ }\textbf {\bibinfo {volume} {88}},\ \bibinfo {pages}
  {125427} (\bibinfo {year} {2013})}\BibitemShut {NoStop}%
\bibitem [{\citenamefont {Weng}\ \emph {et~al.}(2015)\citenamefont {Weng},
  \citenamefont {Fang}, \citenamefont {Fang}, \citenamefont {Bernevig},\ and\
  \citenamefont {Dai}}]{Weng15prx}%
  \BibitemOpen
  \bibfield  {author} {\bibinfo {author} {\bibfnamefont {Hongming}\
  \bibnamefont {Weng}}, \bibinfo {author} {\bibfnamefont {Chen}\ \bibnamefont
  {Fang}}, \bibinfo {author} {\bibfnamefont {Zhong}\ \bibnamefont {Fang}},
  \bibinfo {author} {\bibfnamefont {B.~Andrei}\ \bibnamefont {Bernevig}}, \
  and\ \bibinfo {author} {\bibfnamefont {Xi}~\bibnamefont {Dai}},\ }\bibfield
  {title} {\enquote {\bibinfo {title} {Weyl semimetal phase in
  noncentrosymmetric transition-metal monophosphides},}\ }\href {\doibase
  10.1103/PhysRevX.5.011029} {\bibfield  {journal} {\bibinfo  {journal} {Phys.
  Rev. X}\ }\textbf {\bibinfo {volume} {5}},\ \bibinfo {pages} {011029}
  (\bibinfo {year} {2015})}\BibitemShut {NoStop}%
\bibitem [{\citenamefont {Huang}\ \emph {et~al.}(2015)\citenamefont {Huang},
  \citenamefont {Xu}, \citenamefont {Belopolski}, \citenamefont {Lee},
  \citenamefont {Chang}, \citenamefont {Wang}, \citenamefont {Alidoust},
  \citenamefont {Bian}, \citenamefont {Neupane}, \citenamefont {Zhang} \emph
  {et~al.}}]{huang2015weyl}%
  \BibitemOpen
  \bibfield  {author} {\bibinfo {author} {\bibfnamefont {Shin-Ming}\
  \bibnamefont {Huang}}, \bibinfo {author} {\bibfnamefont {Su-Yang}\
  \bibnamefont {Xu}}, \bibinfo {author} {\bibfnamefont {Ilya}\ \bibnamefont
  {Belopolski}}, \bibinfo {author} {\bibfnamefont {Chi-Cheng}\ \bibnamefont
  {Lee}}, \bibinfo {author} {\bibfnamefont {Guoqing}\ \bibnamefont {Chang}},
  \bibinfo {author} {\bibfnamefont {BaoKai}\ \bibnamefont {Wang}}, \bibinfo
  {author} {\bibfnamefont {Nasser}\ \bibnamefont {Alidoust}}, \bibinfo {author}
  {\bibfnamefont {Guang}\ \bibnamefont {Bian}}, \bibinfo {author}
  {\bibfnamefont {Madhab}\ \bibnamefont {Neupane}}, \bibinfo {author}
  {\bibfnamefont {Chenglong}\ \bibnamefont {Zhang}},  \emph {et~al.},\
  }\bibfield  {title} {\enquote {\bibinfo {title} {A weyl fermion semimetal
  with surface fermi arcs in the transition metal monopnictide taas class},}\
  }\href@noop {} {\bibfield  {journal} {\bibinfo  {journal} {Nature
  communications}\ }\textbf {\bibinfo {volume} {6}},\ \bibinfo {pages} {7373}
  (\bibinfo {year} {2015})}\BibitemShut {NoStop}%
\bibitem [{\citenamefont {Yang}\ \emph {et~al.}(2011)\citenamefont {Yang},
  \citenamefont {Lu},\ and\ \citenamefont {Ran}}]{Yang11prb}%
  \BibitemOpen
  \bibfield  {author} {\bibinfo {author} {\bibfnamefont {Kai-Yu}\ \bibnamefont
  {Yang}}, \bibinfo {author} {\bibfnamefont {Yuan-Ming}\ \bibnamefont {Lu}}, \
  and\ \bibinfo {author} {\bibfnamefont {Ying}\ \bibnamefont {Ran}},\
  }\bibfield  {title} {\enquote {\bibinfo {title} {Quantum hall effects in a
  weyl semimetal: Possible application in pyrochlore iridates},}\ }\href
  {\doibase 10.1103/PhysRevB.84.075129} {\bibfield  {journal} {\bibinfo
  {journal} {Phys. Rev. B}\ }\textbf {\bibinfo {volume} {84}},\ \bibinfo
  {pages} {075129} (\bibinfo {year} {2011})}\BibitemShut {NoStop}%
\bibitem [{\citenamefont {Chan}\ \emph {et~al.}(2017)\citenamefont {Chan},
  \citenamefont {Lindner}, \citenamefont {Refael},\ and\ \citenamefont
  {Lee}}]{chan2017photocurrents}%
  \BibitemOpen
  \bibfield  {author} {\bibinfo {author} {\bibfnamefont {Ching-Kit}\
  \bibnamefont {Chan}}, \bibinfo {author} {\bibfnamefont {Netanel~H.}\
  \bibnamefont {Lindner}}, \bibinfo {author} {\bibfnamefont {Gil}\ \bibnamefont
  {Refael}}, \ and\ \bibinfo {author} {\bibfnamefont {Patrick~A.}\ \bibnamefont
  {Lee}},\ }\bibfield  {title} {\enquote {\bibinfo {title} {Photocurrents in
  weyl semimetals},}\ }\href {\doibase 10.1103/PhysRevB.95.041104} {\bibfield
  {journal} {\bibinfo  {journal} {Phys. Rev. B}\ }\textbf {\bibinfo {volume}
  {95}},\ \bibinfo {pages} {041104} (\bibinfo {year} {2017})}\BibitemShut
  {NoStop}%
\bibitem [{\citenamefont {Osterhoudt}\ \emph {et~al.}(2019)\citenamefont
  {Osterhoudt}, \citenamefont {Diebel}, \citenamefont {Gray}, \citenamefont
  {Yang}, \citenamefont {Stanco}, \citenamefont {Huang}, \citenamefont {Shen},
  \citenamefont {Ni}, \citenamefont {Moll}, \citenamefont {Ran} \emph
  {et~al.}}]{osterhoudt2019colossal}%
  \BibitemOpen
  \bibfield  {author} {\bibinfo {author} {\bibfnamefont {Gavin~B}\ \bibnamefont
  {Osterhoudt}}, \bibinfo {author} {\bibfnamefont {Laura~K}\ \bibnamefont
  {Diebel}}, \bibinfo {author} {\bibfnamefont {Mason~J}\ \bibnamefont {Gray}},
  \bibinfo {author} {\bibfnamefont {Xu}~\bibnamefont {Yang}}, \bibinfo {author}
  {\bibfnamefont {John}\ \bibnamefont {Stanco}}, \bibinfo {author}
  {\bibfnamefont {Xiangwei}\ \bibnamefont {Huang}}, \bibinfo {author}
  {\bibfnamefont {Bing}\ \bibnamefont {Shen}}, \bibinfo {author} {\bibfnamefont
  {Ni}~\bibnamefont {Ni}}, \bibinfo {author} {\bibfnamefont {Philip~JW}\
  \bibnamefont {Moll}}, \bibinfo {author} {\bibfnamefont {Ying}\ \bibnamefont
  {Ran}},  \emph {et~al.},\ }\bibfield  {title} {\enquote {\bibinfo {title}
  {Colossal mid-infrared bulk photovoltaic effect in a type-i weyl
  semimetal},}\ }\href@noop {} {\bibfield  {journal} {\bibinfo  {journal}
  {Nature materials}\ }\textbf {\bibinfo {volume} {18}},\ \bibinfo {pages}
  {471} (\bibinfo {year} {2019})}\BibitemShut {NoStop}%
\bibitem [{\citenamefont {Lv}\ \emph {et~al.}(2015{\natexlab{a}})\citenamefont
  {Lv}, \citenamefont {Weng}, \citenamefont {Fu}, \citenamefont {Wang},
  \citenamefont {Miao}, \citenamefont {Ma}, \citenamefont {Richard},
  \citenamefont {Huang}, \citenamefont {Zhao}, \citenamefont {Chen} \emph
  {et~al.}}]{lv2015experimental}%
  \BibitemOpen
  \bibfield  {author} {\bibinfo {author} {\bibfnamefont {BQ}~\bibnamefont
  {Lv}}, \bibinfo {author} {\bibfnamefont {HM}~\bibnamefont {Weng}}, \bibinfo
  {author} {\bibfnamefont {BB}~\bibnamefont {Fu}}, \bibinfo {author}
  {\bibfnamefont {XP}~\bibnamefont {Wang}}, \bibinfo {author} {\bibfnamefont
  {Hu}~\bibnamefont {Miao}}, \bibinfo {author} {\bibfnamefont {Junzhang}\
  \bibnamefont {Ma}}, \bibinfo {author} {\bibfnamefont {P}~\bibnamefont
  {Richard}}, \bibinfo {author} {\bibfnamefont {XC}~\bibnamefont {Huang}},
  \bibinfo {author} {\bibfnamefont {LX}~\bibnamefont {Zhao}}, \bibinfo {author}
  {\bibfnamefont {GF}~\bibnamefont {Chen}},  \emph {et~al.},\ }\bibfield
  {title} {\enquote {\bibinfo {title} {Experimental discovery of weyl semimetal
  taas},}\ }\href@noop {} {\bibfield  {journal} {\bibinfo  {journal} {Physical
  Review X}\ }\textbf {\bibinfo {volume} {5}},\ \bibinfo {pages} {031013}
  (\bibinfo {year} {2015}{\natexlab{a}})}\BibitemShut {NoStop}%
\bibitem [{\citenamefont {Xu}\ \emph {et~al.}(2015{\natexlab{a}})\citenamefont
  {Xu}, \citenamefont {Belopolski}, \citenamefont {Alidoust}, \citenamefont
  {Neupane}, \citenamefont {Bian}, \citenamefont {Zhang}, \citenamefont
  {Sankar}, \citenamefont {Chang}, \citenamefont {Yuan}, \citenamefont {Lee}
  \emph {et~al.}}]{xu2015discovery}%
  \BibitemOpen
  \bibfield  {author} {\bibinfo {author} {\bibfnamefont {Su-Yang}\ \bibnamefont
  {Xu}}, \bibinfo {author} {\bibfnamefont {Ilya}\ \bibnamefont {Belopolski}},
  \bibinfo {author} {\bibfnamefont {Nasser}\ \bibnamefont {Alidoust}}, \bibinfo
  {author} {\bibfnamefont {Madhab}\ \bibnamefont {Neupane}}, \bibinfo {author}
  {\bibfnamefont {Guang}\ \bibnamefont {Bian}}, \bibinfo {author}
  {\bibfnamefont {Chenglong}\ \bibnamefont {Zhang}}, \bibinfo {author}
  {\bibfnamefont {Raman}\ \bibnamefont {Sankar}}, \bibinfo {author}
  {\bibfnamefont {Guoqing}\ \bibnamefont {Chang}}, \bibinfo {author}
  {\bibfnamefont {Zhujun}\ \bibnamefont {Yuan}}, \bibinfo {author}
  {\bibfnamefont {Chi-Cheng}\ \bibnamefont {Lee}},  \emph {et~al.},\ }\bibfield
   {title} {\enquote {\bibinfo {title} {Discovery of a weyl fermion semimetal
  and topological fermi arcs},}\ }\href@noop {} {\bibfield  {journal} {\bibinfo
   {journal} {Science}\ }\textbf {\bibinfo {volume} {349}},\ \bibinfo {pages}
  {613--617} (\bibinfo {year} {2015}{\natexlab{a}})}\BibitemShut {NoStop}%
\bibitem [{\citenamefont {Xu}\ \emph {et~al.}(2015{\natexlab{b}})\citenamefont
  {Xu}, \citenamefont {Alidoust}, \citenamefont {Belopolski}, \citenamefont
  {Yuan}, \citenamefont {Bian}, \citenamefont {Chang}, \citenamefont {Zheng},
  \citenamefont {Strocov}, \citenamefont {Sanchez}, \citenamefont {Chang} \emph
  {et~al.}}]{xu2015discovery2}%
  \BibitemOpen
  \bibfield  {author} {\bibinfo {author} {\bibfnamefont {Su-Yang}\ \bibnamefont
  {Xu}}, \bibinfo {author} {\bibfnamefont {Nasser}\ \bibnamefont {Alidoust}},
  \bibinfo {author} {\bibfnamefont {Ilya}\ \bibnamefont {Belopolski}}, \bibinfo
  {author} {\bibfnamefont {Zhujun}\ \bibnamefont {Yuan}}, \bibinfo {author}
  {\bibfnamefont {Guang}\ \bibnamefont {Bian}}, \bibinfo {author}
  {\bibfnamefont {Tay-Rong}\ \bibnamefont {Chang}}, \bibinfo {author}
  {\bibfnamefont {Hao}\ \bibnamefont {Zheng}}, \bibinfo {author} {\bibfnamefont
  {Vladimir~N}\ \bibnamefont {Strocov}}, \bibinfo {author} {\bibfnamefont
  {Daniel~S}\ \bibnamefont {Sanchez}}, \bibinfo {author} {\bibfnamefont
  {Guoqing}\ \bibnamefont {Chang}},  \emph {et~al.},\ }\bibfield  {title}
  {\enquote {\bibinfo {title} {Discovery of a weyl fermion state with fermi
  arcs in niobium arsenide},}\ }\href@noop {} {\bibfield  {journal} {\bibinfo
  {journal} {Nature Physics}\ }\textbf {\bibinfo {volume} {11}},\ \bibinfo
  {pages} {748} (\bibinfo {year} {2015}{\natexlab{b}})}\BibitemShut {NoStop}%
\bibitem [{\citenamefont {Xu}\ \emph {et~al.}(2015{\natexlab{c}})\citenamefont
  {Xu}, \citenamefont {Belopolski}, \citenamefont {Sanchez}, \citenamefont
  {Zhang}, \citenamefont {Chang}, \citenamefont {Guo}, \citenamefont {Bian},
  \citenamefont {Yuan}, \citenamefont {Lu}, \citenamefont {Chang} \emph
  {et~al.}}]{xu2015experimental}%
  \BibitemOpen
  \bibfield  {author} {\bibinfo {author} {\bibfnamefont {Su-Yang}\ \bibnamefont
  {Xu}}, \bibinfo {author} {\bibfnamefont {Ilya}\ \bibnamefont {Belopolski}},
  \bibinfo {author} {\bibfnamefont {Daniel~S}\ \bibnamefont {Sanchez}},
  \bibinfo {author} {\bibfnamefont {Chenglong}\ \bibnamefont {Zhang}}, \bibinfo
  {author} {\bibfnamefont {Guoqing}\ \bibnamefont {Chang}}, \bibinfo {author}
  {\bibfnamefont {Cheng}\ \bibnamefont {Guo}}, \bibinfo {author} {\bibfnamefont
  {Guang}\ \bibnamefont {Bian}}, \bibinfo {author} {\bibfnamefont {Zhujun}\
  \bibnamefont {Yuan}}, \bibinfo {author} {\bibfnamefont {Hong}\ \bibnamefont
  {Lu}}, \bibinfo {author} {\bibfnamefont {Tay-Rong}\ \bibnamefont {Chang}},
  \emph {et~al.},\ }\bibfield  {title} {\enquote {\bibinfo {title}
  {Experimental discovery of a topological weyl semimetal state in tap},}\
  }\href@noop {} {\bibfield  {journal} {\bibinfo  {journal} {Science advances}\
  }\textbf {\bibinfo {volume} {1}},\ \bibinfo {pages} {e1501092} (\bibinfo
  {year} {2015}{\natexlab{c}})}\BibitemShut {NoStop}%
\bibitem [{\citenamefont {Xu}\ \emph {et~al.}(2016)\citenamefont {Xu},
  \citenamefont {Weng}, \citenamefont {Lv}, \citenamefont {Matt}, \citenamefont
  {Park}, \citenamefont {Bisti}, \citenamefont {Strocov}, \citenamefont
  {Gawryluk}, \citenamefont {Pomjakushina}, \citenamefont {Conder} \emph
  {et~al.}}]{xu2016observation}%
  \BibitemOpen
  \bibfield  {author} {\bibinfo {author} {\bibfnamefont {Nan}\ \bibnamefont
  {Xu}}, \bibinfo {author} {\bibfnamefont {HM}~\bibnamefont {Weng}}, \bibinfo
  {author} {\bibfnamefont {BQ}~\bibnamefont {Lv}}, \bibinfo {author}
  {\bibfnamefont {Christian~E}\ \bibnamefont {Matt}}, \bibinfo {author}
  {\bibfnamefont {Jihwey}\ \bibnamefont {Park}}, \bibinfo {author}
  {\bibfnamefont {Federico}\ \bibnamefont {Bisti}}, \bibinfo {author}
  {\bibfnamefont {Vladimir~N}\ \bibnamefont {Strocov}}, \bibinfo {author}
  {\bibfnamefont {Dariusz}\ \bibnamefont {Gawryluk}}, \bibinfo {author}
  {\bibfnamefont {Ekaterina}\ \bibnamefont {Pomjakushina}}, \bibinfo {author}
  {\bibfnamefont {Kazimierz}\ \bibnamefont {Conder}},  \emph {et~al.},\
  }\bibfield  {title} {\enquote {\bibinfo {title} {Observation of weyl nodes
  and fermi arcs in tantalum phosphide},}\ }\href@noop {} {\bibfield  {journal}
  {\bibinfo  {journal} {Nature communications}\ }\textbf {\bibinfo {volume}
  {7}},\ \bibinfo {pages} {11006} (\bibinfo {year} {2016})}\BibitemShut
  {NoStop}%
\bibitem [{\citenamefont {Deng}\ \emph {et~al.}(2016)\citenamefont {Deng},
  \citenamefont {Wan}, \citenamefont {Deng}, \citenamefont {Zhang},
  \citenamefont {Ding}, \citenamefont {Wang}, \citenamefont {Yan},
  \citenamefont {Huang}, \citenamefont {Zhang}, \citenamefont {Xu} \emph
  {et~al.}}]{deng2016experimental}%
  \BibitemOpen
  \bibfield  {author} {\bibinfo {author} {\bibfnamefont {Ke}~\bibnamefont
  {Deng}}, \bibinfo {author} {\bibfnamefont {Guoliang}\ \bibnamefont {Wan}},
  \bibinfo {author} {\bibfnamefont {Peng}\ \bibnamefont {Deng}}, \bibinfo
  {author} {\bibfnamefont {Kenan}\ \bibnamefont {Zhang}}, \bibinfo {author}
  {\bibfnamefont {Shijie}\ \bibnamefont {Ding}}, \bibinfo {author}
  {\bibfnamefont {Eryin}\ \bibnamefont {Wang}}, \bibinfo {author}
  {\bibfnamefont {Mingzhe}\ \bibnamefont {Yan}}, \bibinfo {author}
  {\bibfnamefont {Huaqing}\ \bibnamefont {Huang}}, \bibinfo {author}
  {\bibfnamefont {Hongyun}\ \bibnamefont {Zhang}}, \bibinfo {author}
  {\bibfnamefont {Zhilin}\ \bibnamefont {Xu}},  \emph {et~al.},\ }\bibfield
  {title} {\enquote {\bibinfo {title} {Experimental observation of topological
  fermi arcs in type-ii weyl semimetal mote 2},}\ }\href@noop {} {\bibfield
  {journal} {\bibinfo  {journal} {Nature Physics}\ }\textbf {\bibinfo {volume}
  {12}},\ \bibinfo {pages} {1105} (\bibinfo {year} {2016})}\BibitemShut
  {NoStop}%
\bibitem [{\citenamefont {Yang}\ \emph {et~al.}(2015)\citenamefont {Yang},
  \citenamefont {Liu}, \citenamefont {Sun}, \citenamefont {Peng}, \citenamefont
  {Yang}, \citenamefont {Zhang}, \citenamefont {Zhou}, \citenamefont {Zhang},
  \citenamefont {Guo}, \citenamefont {Rahn} \emph {et~al.}}]{yang2015weyl}%
  \BibitemOpen
  \bibfield  {author} {\bibinfo {author} {\bibfnamefont {LX}~\bibnamefont
  {Yang}}, \bibinfo {author} {\bibfnamefont {ZK}~\bibnamefont {Liu}}, \bibinfo
  {author} {\bibfnamefont {Yan}\ \bibnamefont {Sun}}, \bibinfo {author}
  {\bibfnamefont {Han}\ \bibnamefont {Peng}}, \bibinfo {author} {\bibfnamefont
  {HF}~\bibnamefont {Yang}}, \bibinfo {author} {\bibfnamefont {Teng}\
  \bibnamefont {Zhang}}, \bibinfo {author} {\bibfnamefont {Bo}~\bibnamefont
  {Zhou}}, \bibinfo {author} {\bibfnamefont {Yi}~\bibnamefont {Zhang}},
  \bibinfo {author} {\bibfnamefont {YF}~\bibnamefont {Guo}}, \bibinfo {author}
  {\bibfnamefont {Marein}\ \bibnamefont {Rahn}},  \emph {et~al.},\ }\bibfield
  {title} {\enquote {\bibinfo {title} {Weyl semimetal phase in the
  non-centrosymmetric compound taas},}\ }\href@noop {} {\bibfield  {journal}
  {\bibinfo  {journal} {Nature physics}\ }\textbf {\bibinfo {volume} {11}},\
  \bibinfo {pages} {728} (\bibinfo {year} {2015})}\BibitemShut {NoStop}%
\bibitem [{\citenamefont {Huang}\ \emph {et~al.}(2016)\citenamefont {Huang},
  \citenamefont {McCormick}, \citenamefont {Ochi}, \citenamefont {Zhao},
  \citenamefont {Suzuki}, \citenamefont {Arita}, \citenamefont {Wu},
  \citenamefont {Mou}, \citenamefont {Cao}, \citenamefont {Yan} \emph
  {et~al.}}]{huang2016spectroscopic}%
  \BibitemOpen
  \bibfield  {author} {\bibinfo {author} {\bibfnamefont {Lunan}\ \bibnamefont
  {Huang}}, \bibinfo {author} {\bibfnamefont {Timothy~M}\ \bibnamefont
  {McCormick}}, \bibinfo {author} {\bibfnamefont {Masayuki}\ \bibnamefont
  {Ochi}}, \bibinfo {author} {\bibfnamefont {Zhiying}\ \bibnamefont {Zhao}},
  \bibinfo {author} {\bibfnamefont {Michi-To}\ \bibnamefont {Suzuki}}, \bibinfo
  {author} {\bibfnamefont {Ryotaro}\ \bibnamefont {Arita}}, \bibinfo {author}
  {\bibfnamefont {Yun}\ \bibnamefont {Wu}}, \bibinfo {author} {\bibfnamefont
  {Daixiang}\ \bibnamefont {Mou}}, \bibinfo {author} {\bibfnamefont {Huibo}\
  \bibnamefont {Cao}}, \bibinfo {author} {\bibfnamefont {Jiaqiang}\
  \bibnamefont {Yan}},  \emph {et~al.},\ }\bibfield  {title} {\enquote
  {\bibinfo {title} {Spectroscopic evidence for a type ii weyl semimetallic
  state in mote 2},}\ }\href@noop {} {\bibfield  {journal} {\bibinfo  {journal}
  {Nature materials}\ }\textbf {\bibinfo {volume} {15}},\ \bibinfo {pages}
  {1155} (\bibinfo {year} {2016})}\BibitemShut {NoStop}%
\bibitem [{\citenamefont {Tamai}\ \emph {et~al.}(2016)\citenamefont {Tamai},
  \citenamefont {Wu}, \citenamefont {Cucchi}, \citenamefont {Bruno},
  \citenamefont {Ricc{\`o}}, \citenamefont {Kim}, \citenamefont {Hoesch},
  \citenamefont {Barreteau}, \citenamefont {Giannini}, \citenamefont {Besnard}
  \emph {et~al.}}]{tamai2016fermi}%
  \BibitemOpen
  \bibfield  {author} {\bibinfo {author} {\bibfnamefont {Anna}\ \bibnamefont
  {Tamai}}, \bibinfo {author} {\bibfnamefont {QS}~\bibnamefont {Wu}}, \bibinfo
  {author} {\bibfnamefont {Ir{\`e}ne}\ \bibnamefont {Cucchi}}, \bibinfo
  {author} {\bibfnamefont {Flavio~Yair}\ \bibnamefont {Bruno}}, \bibinfo
  {author} {\bibfnamefont {Sara}\ \bibnamefont {Ricc{\`o}}}, \bibinfo {author}
  {\bibfnamefont {TK}~\bibnamefont {Kim}}, \bibinfo {author} {\bibfnamefont
  {M}~\bibnamefont {Hoesch}}, \bibinfo {author} {\bibfnamefont {C{\'e}line}\
  \bibnamefont {Barreteau}}, \bibinfo {author} {\bibfnamefont {Enrico}\
  \bibnamefont {Giannini}}, \bibinfo {author} {\bibfnamefont {C{\'e}line}\
  \bibnamefont {Besnard}},  \emph {et~al.},\ }\bibfield  {title} {\enquote
  {\bibinfo {title} {Fermi arcs and their topological character in the
  candidate type-ii weyl semimetal mote 2},}\ }\href@noop {} {\bibfield
  {journal} {\bibinfo  {journal} {Physical Review X}\ }\textbf {\bibinfo
  {volume} {6}},\ \bibinfo {pages} {031021} (\bibinfo {year}
  {2016})}\BibitemShut {NoStop}%
\bibitem [{\citenamefont {Jiang}\ \emph {et~al.}(2017)\citenamefont {Jiang},
  \citenamefont {Liu}, \citenamefont {Sun}, \citenamefont {Yang}, \citenamefont
  {Rajamathi}, \citenamefont {Qi}, \citenamefont {Yang}, \citenamefont {Chen},
  \citenamefont {Peng}, \citenamefont {Hwang} \emph
  {et~al.}}]{jiang2017signature}%
  \BibitemOpen
  \bibfield  {author} {\bibinfo {author} {\bibfnamefont {Juan}\ \bibnamefont
  {Jiang}}, \bibinfo {author} {\bibfnamefont {ZK}~\bibnamefont {Liu}}, \bibinfo
  {author} {\bibfnamefont {Y}~\bibnamefont {Sun}}, \bibinfo {author}
  {\bibfnamefont {HF}~\bibnamefont {Yang}}, \bibinfo {author} {\bibfnamefont
  {CR}~\bibnamefont {Rajamathi}}, \bibinfo {author} {\bibfnamefont
  {YP}~\bibnamefont {Qi}}, \bibinfo {author} {\bibfnamefont {LX}~\bibnamefont
  {Yang}}, \bibinfo {author} {\bibfnamefont {C}~\bibnamefont {Chen}}, \bibinfo
  {author} {\bibfnamefont {H}~\bibnamefont {Peng}}, \bibinfo {author}
  {\bibfnamefont {CC}~\bibnamefont {Hwang}},  \emph {et~al.},\ }\bibfield
  {title} {\enquote {\bibinfo {title} {Signature of type-ii weyl semimetal
  phase in mote 2},}\ }\href@noop {} {\bibfield  {journal} {\bibinfo  {journal}
  {Nature communications}\ }\textbf {\bibinfo {volume} {8}},\ \bibinfo {pages}
  {13973} (\bibinfo {year} {2017})}\BibitemShut {NoStop}%
\bibitem [{\citenamefont {Belopolski}\ \emph {et~al.}(2016)\citenamefont
  {Belopolski}, \citenamefont {Sanchez}, \citenamefont {Ishida}, \citenamefont
  {Pan}, \citenamefont {Yu}, \citenamefont {Xu}, \citenamefont {Chang},
  \citenamefont {Chang}, \citenamefont {Zheng}, \citenamefont {Alidoust} \emph
  {et~al.}}]{belopolski2016discovery}%
  \BibitemOpen
  \bibfield  {author} {\bibinfo {author} {\bibfnamefont {Ilya}\ \bibnamefont
  {Belopolski}}, \bibinfo {author} {\bibfnamefont {Daniel~S}\ \bibnamefont
  {Sanchez}}, \bibinfo {author} {\bibfnamefont {Yukiaki}\ \bibnamefont
  {Ishida}}, \bibinfo {author} {\bibfnamefont {Xingchen}\ \bibnamefont {Pan}},
  \bibinfo {author} {\bibfnamefont {Peng}\ \bibnamefont {Yu}}, \bibinfo
  {author} {\bibfnamefont {Su-Yang}\ \bibnamefont {Xu}}, \bibinfo {author}
  {\bibfnamefont {Guoqing}\ \bibnamefont {Chang}}, \bibinfo {author}
  {\bibfnamefont {Tay-Rong}\ \bibnamefont {Chang}}, \bibinfo {author}
  {\bibfnamefont {Hao}\ \bibnamefont {Zheng}}, \bibinfo {author} {\bibfnamefont
  {Nasser}\ \bibnamefont {Alidoust}},  \emph {et~al.},\ }\bibfield  {title}
  {\enquote {\bibinfo {title} {Discovery of a new type of topological weyl
  fermion semimetal state in mo x w 1- x te 2},}\ }\href@noop {} {\bibfield
  {journal} {\bibinfo  {journal} {Nature communications}\ }\textbf {\bibinfo
  {volume} {7}},\ \bibinfo {pages} {13643} (\bibinfo {year}
  {2016})}\BibitemShut {NoStop}%
\bibitem [{\citenamefont {Lv}\ \emph {et~al.}(2015{\natexlab{b}})\citenamefont
  {Lv}, \citenamefont {Xu}, \citenamefont {Weng}, \citenamefont {Ma},
  \citenamefont {Richard}, \citenamefont {Huang}, \citenamefont {Zhao},
  \citenamefont {Chen}, \citenamefont {Matt}, \citenamefont {Bisti} \emph
  {et~al.}}]{lv2015observation}%
  \BibitemOpen
  \bibfield  {author} {\bibinfo {author} {\bibfnamefont {BQ}~\bibnamefont
  {Lv}}, \bibinfo {author} {\bibfnamefont {N}~\bibnamefont {Xu}}, \bibinfo
  {author} {\bibfnamefont {HM}~\bibnamefont {Weng}}, \bibinfo {author}
  {\bibfnamefont {JZ}~\bibnamefont {Ma}}, \bibinfo {author} {\bibfnamefont
  {P}~\bibnamefont {Richard}}, \bibinfo {author} {\bibfnamefont
  {XC}~\bibnamefont {Huang}}, \bibinfo {author} {\bibfnamefont
  {LX}~\bibnamefont {Zhao}}, \bibinfo {author} {\bibfnamefont {GF}~\bibnamefont
  {Chen}}, \bibinfo {author} {\bibfnamefont {CE}~\bibnamefont {Matt}}, \bibinfo
  {author} {\bibfnamefont {F}~\bibnamefont {Bisti}},  \emph {et~al.},\
  }\bibfield  {title} {\enquote {\bibinfo {title} {Observation of weyl nodes in
  taas},}\ }\href@noop {} {\bibfield  {journal} {\bibinfo  {journal} {Nature
  Physics}\ }\textbf {\bibinfo {volume} {11}},\ \bibinfo {pages} {724}
  (\bibinfo {year} {2015}{\natexlab{b}})}\BibitemShut {NoStop}%
\bibitem [{\citenamefont {Chen}\ \emph
  {et~al.}(2019{\natexlab{b}})\citenamefont {Chen}, \citenamefont {Chen},\ and\
  \citenamefont {Zilberberg}}]{chen19arxiv}%
  \BibitemOpen
  \bibfield  {author} {\bibinfo {author} {\bibfnamefont {Guangze}\ \bibnamefont
  {Chen}}, \bibinfo {author} {\bibfnamefont {Wei}\ \bibnamefont {Chen}}, \ and\
  \bibinfo {author} {\bibfnamefont {Oded}\ \bibnamefont {Zilberberg}},\
  }\bibfield  {title} {\enquote {\bibinfo {title} {Negative refraction in fermi
  arc surface states of weyl semimetals},}\ }\href@noop {} {\bibfield
  {journal} {\bibinfo  {journal} {in preparation}\ } (\bibinfo {year}
  {2019}{\natexlab{b}})}\BibitemShut {NoStop}%
\bibitem [{\citenamefont {He}\ \emph {et~al.}(2018)\citenamefont {He},
  \citenamefont {Qiu}, \citenamefont {Ye}, \citenamefont {Cai}, \citenamefont
  {Fan}, \citenamefont {Ke}, \citenamefont {Zhang},\ and\ \citenamefont
  {Liu}}]{he2018topological}%
  \BibitemOpen
  \bibfield  {author} {\bibinfo {author} {\bibfnamefont {Hailong}\ \bibnamefont
  {He}}, \bibinfo {author} {\bibfnamefont {Chunyin}\ \bibnamefont {Qiu}},
  \bibinfo {author} {\bibfnamefont {Liping}\ \bibnamefont {Ye}}, \bibinfo
  {author} {\bibfnamefont {Xiangxi}\ \bibnamefont {Cai}}, \bibinfo {author}
  {\bibfnamefont {Xiying}\ \bibnamefont {Fan}}, \bibinfo {author}
  {\bibfnamefont {Manzhu}\ \bibnamefont {Ke}}, \bibinfo {author} {\bibfnamefont
  {Fan}\ \bibnamefont {Zhang}}, \ and\ \bibinfo {author} {\bibfnamefont
  {Zhengyou}\ \bibnamefont {Liu}},\ }\bibfield  {title} {\enquote {\bibinfo
  {title} {Topological negative refraction of surface acoustic waves in a weyl
  phononic crystal},}\ }\href@noop {} {\bibfield  {journal} {\bibinfo
  {journal} {Nature}\ }\textbf {\bibinfo {volume} {560}},\ \bibinfo {pages}
  {61} (\bibinfo {year} {2018})}\BibitemShut {NoStop}%
\bibitem [{\citenamefont {Ruan}\ \emph
  {et~al.}(2016{\natexlab{a}})\citenamefont {Ruan}, \citenamefont {Jian},
  \citenamefont {Yao}, \citenamefont {Zhang}, \citenamefont {Zhang},\ and\
  \citenamefont {Xing}}]{Ruan16nc}%
  \BibitemOpen
  \bibfield  {author} {\bibinfo {author} {\bibfnamefont {Jiawei}\ \bibnamefont
  {Ruan}}, \bibinfo {author} {\bibfnamefont {Shao-Kai}\ \bibnamefont {Jian}},
  \bibinfo {author} {\bibfnamefont {Hong}\ \bibnamefont {Yao}}, \bibinfo
  {author} {\bibfnamefont {Haijun}\ \bibnamefont {Zhang}}, \bibinfo {author}
  {\bibfnamefont {Shou-Cheng}\ \bibnamefont {Zhang}}, \ and\ \bibinfo {author}
  {\bibfnamefont {Dingyu}\ \bibnamefont {Xing}},\ }\bibfield  {title} {\enquote
  {\bibinfo {title} {Symmetry-protected ideal weyl semimetal in hgte-class
  materials},}\ }\href@noop {} {\bibfield  {journal} {\bibinfo  {journal}
  {Nature communications}\ }\textbf {\bibinfo {volume} {7}},\ \bibinfo {pages}
  {11136} (\bibinfo {year} {2016}{\natexlab{a}})}\BibitemShut {NoStop}%
\bibitem [{\citenamefont {Ruan}\ \emph
  {et~al.}(2016{\natexlab{b}})\citenamefont {Ruan}, \citenamefont {Jian},
  \citenamefont {Zhang}, \citenamefont {Yao}, \citenamefont {Zhang},
  \citenamefont {Zhang},\ and\ \citenamefont {Xing}}]{Ruan16prl}%
  \BibitemOpen
  \bibfield  {author} {\bibinfo {author} {\bibfnamefont {Jiawei}\ \bibnamefont
  {Ruan}}, \bibinfo {author} {\bibfnamefont {Shao-Kai}\ \bibnamefont {Jian}},
  \bibinfo {author} {\bibfnamefont {Dongqin}\ \bibnamefont {Zhang}}, \bibinfo
  {author} {\bibfnamefont {Hong}\ \bibnamefont {Yao}}, \bibinfo {author}
  {\bibfnamefont {Haijun}\ \bibnamefont {Zhang}}, \bibinfo {author}
  {\bibfnamefont {Shou-Cheng}\ \bibnamefont {Zhang}}, \ and\ \bibinfo {author}
  {\bibfnamefont {Dingyu}\ \bibnamefont {Xing}},\ }\bibfield  {title} {\enquote
  {\bibinfo {title} {Ideal weyl semimetals in the chalcopyrites
  ${\mathrm{cutlse}}_{2}$, ${\mathrm{agtlte}}_{2}$, ${\mathrm{autlte}}_{2}$,
  and ${\mathrm{znpbas}}_{2}$},}\ }\href {\doibase
  10.1103/PhysRevLett.116.226801} {\bibfield  {journal} {\bibinfo  {journal}
  {Phys. Rev. Lett.}\ }\textbf {\bibinfo {volume} {116}},\ \bibinfo {pages}
  {226801} (\bibinfo {year} {2016}{\natexlab{b}})}\BibitemShut {NoStop}%
\bibitem [{\citenamefont {Chen}\ \emph {et~al.}(2013)\citenamefont {Chen},
  \citenamefont {Jiang}, \citenamefont {Shen}, \citenamefont {Sheng},
  \citenamefont {Wang},\ and\ \citenamefont {Xing}}]{chen2013specular}%
  \BibitemOpen
  \bibfield  {author} {\bibinfo {author} {\bibfnamefont {Wei}\ \bibnamefont
  {Chen}}, \bibinfo {author} {\bibfnamefont {Liang}\ \bibnamefont {Jiang}},
  \bibinfo {author} {\bibfnamefont {R}~\bibnamefont {Shen}}, \bibinfo {author}
  {\bibfnamefont {L}~\bibnamefont {Sheng}}, \bibinfo {author} {\bibfnamefont
  {BG}~\bibnamefont {Wang}}, \ and\ \bibinfo {author} {\bibfnamefont
  {DY}~\bibnamefont {Xing}},\ }\bibfield  {title} {\enquote {\bibinfo {title}
  {Specular andreev reflection in inversion-symmetric weyl semimetals},}\
  }\href@noop {} {\bibfield  {journal} {\bibinfo  {journal} {EPL (Europhysics
  Letters)}\ }\textbf {\bibinfo {volume} {103}},\ \bibinfo {pages} {27006}
  (\bibinfo {year} {2013})}\BibitemShut {NoStop}%
\bibitem [{ano()}]{another_fn}%
  \BibitemOpen
  \href@noop {} {}\bibinfo {note} {The switch-off angle $\theta_0$ is given by
  $\theta_0=\arctan\frac{v_0}{\sqrt{2}k_0d}$ for the effective Hamiltonian
  Eq.\eqref{surface_H}}\BibitemShut {NoStop}%
\bibitem [{foo()}]{footnote}%
  \BibitemOpen
  \href@noop {} {}\bibinfo {note} {For curved Fermi arcs and small $\theta$,
  there exist reflection processes in addition to the negative refraction.
  Here, the reflection does not change $v_z$, and we calculate $\bar{v}_z$ by
  $\bar{v}_z=\int^{k^1_x}_{-k^0_x}\frac{dk_x}{2k^0_x}\frac{v^{\text{I}}_zv^{\text{II}}_x+v^{\text{II}}_zv^{\text{I}}_x}{v^{\text{I}}_x+v^{\text{II}}_x}+\int_{k^1_x}^{k^0_x}\frac{dk_x}{2k^0_x}v^{\text{I}}_z$,
  where $k^1_x=-k_0\sin\theta/\sqrt{1+\sin^2\theta}+\hbar v_0\cos\theta/d$
  seperates the Fermi arc into two segments which lead to negative refraction
  and reflection, respectively.}\BibitemShut {Stop}%
\bibitem [{\citenamefont {Groth}\ \emph {et~al.}(2014)\citenamefont {Groth},
  \citenamefont {Wimmer}, \citenamefont {Akhmerov},\ and\ \citenamefont
  {Waintal}}]{groth2014kwant}%
  \BibitemOpen
  \bibfield  {author} {\bibinfo {author} {\bibfnamefont {Christoph~W}\
  \bibnamefont {Groth}}, \bibinfo {author} {\bibfnamefont {Michael}\
  \bibnamefont {Wimmer}}, \bibinfo {author} {\bibfnamefont {Anton~R}\
  \bibnamefont {Akhmerov}}, \ and\ \bibinfo {author} {\bibfnamefont {Xavier}\
  \bibnamefont {Waintal}},\ }\bibfield  {title} {\enquote {\bibinfo {title}
  {Kwant: a software package for quantum transport},}\ }\href@noop {}
  {\bibfield  {journal} {\bibinfo  {journal} {New Journal of Physics}\ }\textbf
  {\bibinfo {volume} {16}},\ \bibinfo {pages} {063065} (\bibinfo {year}
  {2014})}\BibitemShut {NoStop}%
\bibitem [{\citenamefont {Yang}\ \emph {et~al.}(2019)\citenamefont {Yang},
  \citenamefont {Yang}, \citenamefont {Liu}, \citenamefont {Sun}, \citenamefont
  {Chen}, \citenamefont {Peng}, \citenamefont {Schmidt}, \citenamefont
  {Prabhakaran}, \citenamefont {Bernevig}, \citenamefont {Felser} \emph
  {et~al.}}]{yang2019topological}%
  \BibitemOpen
  \bibfield  {author} {\bibinfo {author} {\bibfnamefont {HF}~\bibnamefont
  {Yang}}, \bibinfo {author} {\bibfnamefont {LX}~\bibnamefont {Yang}}, \bibinfo
  {author} {\bibfnamefont {ZK}~\bibnamefont {Liu}}, \bibinfo {author}
  {\bibfnamefont {Y}~\bibnamefont {Sun}}, \bibinfo {author} {\bibfnamefont
  {C}~\bibnamefont {Chen}}, \bibinfo {author} {\bibfnamefont {H}~\bibnamefont
  {Peng}}, \bibinfo {author} {\bibfnamefont {M}~\bibnamefont {Schmidt}},
  \bibinfo {author} {\bibfnamefont {D}~\bibnamefont {Prabhakaran}}, \bibinfo
  {author} {\bibfnamefont {BA}~\bibnamefont {Bernevig}}, \bibinfo {author}
  {\bibfnamefont {C}~\bibnamefont {Felser}},  \emph {et~al.},\ }\bibfield
  {title} {\enquote {\bibinfo {title} {Topological lifshitz transitions and
  fermi arc manipulation in weyl semimetal nbas},}\ }\href@noop {} {\bibfield
  {journal} {\bibinfo  {journal} {Nature communications}\ }\textbf {\bibinfo
  {volume} {10}},\ \bibinfo {pages} {1--7} (\bibinfo {year}
  {2019})}\BibitemShut {NoStop}%
\bibitem [{\citenamefont {Morali}\ \emph {et~al.}(2019)\citenamefont {Morali},
  \citenamefont {Batabyal}, \citenamefont {Nag}, \citenamefont {Liu},
  \citenamefont {Xu}, \citenamefont {Sun}, \citenamefont {Yan}, \citenamefont
  {Felser}, \citenamefont {Avraham},\ and\ \citenamefont
  {Beidenkopf}}]{morali2019fermi}%
  \BibitemOpen
  \bibfield  {author} {\bibinfo {author} {\bibfnamefont {Noam}\ \bibnamefont
  {Morali}}, \bibinfo {author} {\bibfnamefont {Rajib}\ \bibnamefont
  {Batabyal}}, \bibinfo {author} {\bibfnamefont {Pranab~Kumar}\ \bibnamefont
  {Nag}}, \bibinfo {author} {\bibfnamefont {Enke}\ \bibnamefont {Liu}},
  \bibinfo {author} {\bibfnamefont {Qiunan}\ \bibnamefont {Xu}}, \bibinfo
  {author} {\bibfnamefont {Yan}\ \bibnamefont {Sun}}, \bibinfo {author}
  {\bibfnamefont {Binghai}\ \bibnamefont {Yan}}, \bibinfo {author}
  {\bibfnamefont {Claudia}\ \bibnamefont {Felser}}, \bibinfo {author}
  {\bibfnamefont {Nurit}\ \bibnamefont {Avraham}}, \ and\ \bibinfo {author}
  {\bibfnamefont {Haim}\ \bibnamefont {Beidenkopf}},\ }\bibfield  {title}
  {\enquote {\bibinfo {title} {Fermi-arc diversity on surface terminations of
  the magnetic weyl semimetal co3sn2s2},}\ }\href@noop {} {\bibfield  {journal}
  {\bibinfo  {journal} {arXiv preprint arXiv:1903.00509}\ } (\bibinfo {year}
  {2019})}\BibitemShut {NoStop}%
\bibitem [{\citenamefont {Li}\ \emph {et~al.}(2019)\citenamefont {Li},
  \citenamefont {Li}, \citenamefont {Du}, \citenamefont {Wang}, \citenamefont
  {Gu}, \citenamefont {Zhang}, \citenamefont {He}, \citenamefont {Duan},\ and\
  \citenamefont {Xu}}]{MnBi2Te4}%
  \BibitemOpen
  \bibfield  {author} {\bibinfo {author} {\bibfnamefont {Jiaheng}\ \bibnamefont
  {Li}}, \bibinfo {author} {\bibfnamefont {Yang}\ \bibnamefont {Li}}, \bibinfo
  {author} {\bibfnamefont {Shiqiao}\ \bibnamefont {Du}}, \bibinfo {author}
  {\bibfnamefont {Zun}\ \bibnamefont {Wang}}, \bibinfo {author} {\bibfnamefont
  {Bing-Lin}\ \bibnamefont {Gu}}, \bibinfo {author} {\bibfnamefont
  {Shou-Cheng}\ \bibnamefont {Zhang}}, \bibinfo {author} {\bibfnamefont
  {Ke}~\bibnamefont {He}}, \bibinfo {author} {\bibfnamefont {Wenhui}\
  \bibnamefont {Duan}}, \ and\ \bibinfo {author} {\bibfnamefont {Yong}\
  \bibnamefont {Xu}},\ }\bibfield  {title} {\enquote {\bibinfo {title}
  {Intrinsic magnetic topological insulators in van der waals layered
  mnbi2te4-family materials},}\ }\href@noop {} {\bibfield  {journal} {\bibinfo
  {journal} {Science Advances}\ }\textbf {\bibinfo {volume} {5}},\ \bibinfo
  {pages} {eaaw5685} (\bibinfo {year} {2019})}\BibitemShut {NoStop}%
\bibitem [{\citenamefont {Wang}\ \emph {et~al.}(2019)\citenamefont {Wang},
  \citenamefont {Jo}, \citenamefont {Kuthanazhi}, \citenamefont {Wu},
  \citenamefont {McQueeney}, \citenamefont {Kaminski},\ and\ \citenamefont
  {Canfield}}]{wang19prb}%
  \BibitemOpen
  \bibfield  {author} {\bibinfo {author} {\bibfnamefont {Lin-Lin}\ \bibnamefont
  {Wang}}, \bibinfo {author} {\bibfnamefont {Na~Hyun}\ \bibnamefont {Jo}},
  \bibinfo {author} {\bibfnamefont {Brinda}\ \bibnamefont {Kuthanazhi}},
  \bibinfo {author} {\bibfnamefont {Yun}\ \bibnamefont {Wu}}, \bibinfo {author}
  {\bibfnamefont {Robert~J.}\ \bibnamefont {McQueeney}}, \bibinfo {author}
  {\bibfnamefont {Adam}\ \bibnamefont {Kaminski}}, \ and\ \bibinfo {author}
  {\bibfnamefont {Paul~C.}\ \bibnamefont {Canfield}},\ }\bibfield  {title}
  {\enquote {\bibinfo {title} {Single pair of weyl fermions in the
  half-metallic semimetal
  $\mathrm{EuC}{\mathrm{d}}_{2}\mathrm{A}{\mathrm{s}}_{2}$},}\ }\href {\doibase
  10.1103/PhysRevB.99.245147} {\bibfield  {journal} {\bibinfo  {journal} {Phys.
  Rev. B}\ }\textbf {\bibinfo {volume} {99}},\ \bibinfo {pages} {245147}
  (\bibinfo {year} {2019})}\BibitemShut {NoStop}%
\bibitem [{\citenamefont {Soh}\ \emph {et~al.}(2019)\citenamefont {Soh},
  \citenamefont {de~Juan}, \citenamefont {Vergniory}, \citenamefont
  {Schr{\"o}ter}, \citenamefont {Rahn}, \citenamefont {Yan}, \citenamefont
  {Bristow}, \citenamefont {Reiss}, \citenamefont {Blandy}, \citenamefont {Guo}
  \emph {et~al.}}]{soh2019ideal}%
  \BibitemOpen
  \bibfield  {author} {\bibinfo {author} {\bibfnamefont {J-R}\ \bibnamefont
  {Soh}}, \bibinfo {author} {\bibfnamefont {F}~\bibnamefont {de~Juan}},
  \bibinfo {author} {\bibfnamefont {MG}~\bibnamefont {Vergniory}}, \bibinfo
  {author} {\bibfnamefont {NBM}\ \bibnamefont {Schr{\"o}ter}}, \bibinfo
  {author} {\bibfnamefont {MC}~\bibnamefont {Rahn}}, \bibinfo {author}
  {\bibfnamefont {DY}~\bibnamefont {Yan}}, \bibinfo {author} {\bibfnamefont
  {M}~\bibnamefont {Bristow}}, \bibinfo {author} {\bibfnamefont
  {PA}~\bibnamefont {Reiss}}, \bibinfo {author} {\bibfnamefont
  {JN}~\bibnamefont {Blandy}}, \bibinfo {author} {\bibfnamefont
  {YF}~\bibnamefont {Guo}},  \emph {et~al.},\ }\bibfield  {title} {\enquote
  {\bibinfo {title} {An ideal weyl semimetal induced by magnetic exchange},}\
  }\href@noop {} {\bibfield  {journal} {\bibinfo  {journal} {arXiv preprint
  arXiv:1901.10022}\ } (\bibinfo {year} {2019})}\BibitemShut {NoStop}%
\end{thebibliography}
%

\end{document}